\documentclass[final, 3p,twocolumn]{elsarticle}

\usepackage{lineno}

\usepackage{amsmath}
\usepackage{algorithmic}
\usepackage{algorithm2e}
\usepackage{multirow}

\usepackage{pifont}    
\usepackage{natbib}    
\usepackage{geometry}  
\usepackage{graphicx}  
\usepackage{txfonts}   
\usepackage{hyperref}  

\usepackage{lipsum}
\usepackage[figuresright]{rotating}
\graphicspath{{figures/}}
\usepackage[utf8]{inputenc}
\usepackage{graphicx}
\usepackage{soul}
\usepackage[caption=false,font=footnotesize]{subfig}
\usepackage{epstopdf}
\graphicspath{{figures/}}
\usepackage{natbib}    
\usepackage{gensymb}
\usepackage{booktabs}
\usepackage{multirow}
\usepackage[official]{eurosym} 

\usepackage{enumitem}
\usepackage{upgreek}
\usepackage{textcomp}

\usepackage{listings}
\usepackage{gensymb}    

\newcommand{\figref}[1]{Figure~\ref{fig:#1}}
\newcommand{\figsref}[2]{Figures~\ref{fig:#1}~and~\ref{fig:#2}}
\newcommand{\figtoref}[2]{Figures~\ref{fig:#1}~through~\ref{fig:#2}}

\newcommand{\tblref}[1]{Table~\ref{tbl:#1}}

\newcommand{\eref}[1]{Eq.~\ref{eqn:#1}}

\newcommand{\secref}[1]{Section~\ref{sec:#1}}
\newcommand{\subsec}[1]{Section~\ref{subsec:#1}}
\newcommand{\subsubsec}[1]{Section~\ref{subsubsec:#1}}

\newcommand{\etal}{\mbox{\emph{et al.}}}

\newcommand{\ignore}[1]{} 

\emergencystretch=\maxdimen
\begin{document}

\title{Modeling methodology for the accurate and prompt prediction of symptomatic events in chronic diseases}

\author[ucm,ccs]{Josué Pagán\corref{cor1}}
\ead{jpagan@ucm.es}
\author[ucm]{José L. Risco-Martín}
\ead{jlrisco@ucm.es}
\author[lsi,ccs]{José M. Moya}
\ead{josem@die.upm.es}
\author[ucm]{José L. Ayala}
\ead{jayala@ucm.es}

\cortext[cor1]{Corresponding author}
\address[ucm]{Dpt. of Computer Architecture and Automation, Complutense University of Madrid, Madrid 28040, Spain}
\address[ccs]{CCS-Center for Computational Simulation, Campus de Montegancedo UPM, Boadilla del Monte 28660, Spain}
\address[lsi]{LSI-Integrated Systems Laboratory, Technical University of Madrid, Madrid
  28040, Spain}

\begin{abstract}
Prediction of symptomatic crises in chronic diseases allows to take
decisions before the symptoms occur, such as the intake of drugs to
avoid the symptoms or the activation of medical alarms. The prediction
horizon is in this case an important parameter in order to fulfill the
pharmacokinetics of medications, or the time response of medical
services. This paper presents a study about the prediction limits of a
chronic disease with symptomatic crises: the migraine. For that
purpose, this work develops a methodology to build predictive migraine
models and to improve these predictions beyond the limits of the
initial models. The maximum prediction horizon is analyzed, and its
dependency on the selected features is studied. A strategy for model
selection is proposed to tackle the trade off between conservative but
robust predictive models, with respect to less accurate predictions
with higher horizons. The obtained results show a prediction horizon
close to 40 minutes, which is in the time range of the drug
pharmacokinetics. Experiments have been performed in a realistic
scenario where input data have been acquired in an ambulatory clinical
study by the deployment of a non-intrusive Wireless Body Sensor
Network. Our results provide an effective methodology for the
selection of the future horizon in the development of prediction
algorithms for diseases experiencing symptomatic crises.

\end{abstract}


\maketitle


\section{Introduction}
\label{sec:introduction}

Currently, there exists a growing interest in the use of Wireless Body
Sensor Networks (WBSNs) as an effective mechanism to monitor biometric
signals. These networks are good candidates for the monitorization of
chronic diseases because they are fully portable and non
intrusive. The monitorization process enables the study of diseases,
and also the prediction of critical events related to the disease
during the monitorization period. These networks have become more
popular with the development of high-performance embedded
architectures and the improvement of their battery life. As pointed
out in~\cite{schwiebert2001research}, many applications and
possibilities emerge in areas such as healthcare for the elderly,
remote medical diagnosis, disease alarm
notifications~\cite{alemdar2010wireless} or mobile applications for
sport training~\cite{kugler2011mobile}. The deployment of these WBSNs
involves the management of large medical databases, and the
development of various processing techniques. For example, data mining
techniques~\cite{banaee2013data} applied to data acquired by wearable
systems allow the detection and classification of epileptic seizures,
as Heldberg~\etal~do in~\cite{heldberg2015using}.

Algorithms for modeling and prediction have also been proposed in some
medical areas. There are many pathologies that can benefit from
predictions, such as those presenting symptomatic crises. A symptomatic
crisis is defined as the manifestation of the symptoms of a
disease. Many diseases present symptomatic crises, like strokes,
epileptic seizures, migraines, psychiatric pathologies or even
digestive pathologies. In some cases, prediction of a symptomatic
crisis is crucial for the patient---for example, prediction of heart
attack in cardiovascular diseases~\cite{kutur2012improved}. 

Many chronic diseases with symptomatic crises exhibit changes in the
biosignals regulated by the Autonomic Nervous System (ANS). Some
examples of diseases with affection in the ANS are multiple
sclerosis~\cite{racosta2015autonomic}, Parkinson
disease~\cite{kaufmann2013autonomic} or cluster headaches and
migraines~\cite{boiardi1988impaired}.

ANS controls the hemodynamic signals like body temperature,
electrodermal activity or heart rate. These signals are easily
monitorized in an ambulatory and non-intrusive way, such as ECG or the
body temperature~\cite{babusiak2011eeg}. We understand as ambulatory
monitorization the process that does not require to be conducted in
hospital, i.e., the patients can continue their normal activities. For
the easy and comfortable monitorization of these signals, we can
deploy WBSNs.

The aforementioned predictions of symptomatic crises are sensitive to
the prediction horizon (i.e. the time between declaration of an
hypothetical event and the event itself). Time response between
prediction and the event is a critical period to take decisions, such
as activating a medical alarm or notifying the intake of a drug.

The prediction horizon is a critical parameter. This paper presents
a study of the effectiveness of prediction in the detection of a
symptomatic crisis. Additionally, we will present how our study can be
applied in a real case of a chronic disease, the prediction of
migraines. This case study has been selected because of the
complexity of the problem in terms of modeling and variable selection,
as well as its social-economical impact.

It is known that several hemodynamic variables change regulated by ANS
when a migraine occurs. ANS regulates hemodynamic variables such as
the heart and respiratory rate or sweating and vasomotor
activity. This also happens in migraines. Some previous works on
migraine treatment have demonstrated that, with the usage of
domperidona and naratriptan, the earlier the intake of the medication,
the more effective is. Goadsby~\etal~show some results
in~\cite{goadsby2008early}. There are studies about the usage of other
triptans with a shorter time of actuation. In the same line,
in~\cite{hu2002treatment} it is shown that the pharmacokinetics of
specific migraine treatments, such as rizatriptan or sumatriptan, can
abort migraines in 30 and 10 minutes respectively. Therefore, any
prediction of the migraine crisis would be extremely useful to avoid
the pain before its onset, as the pharmacological treatment already
exists.

Some of the migraineurs (defined in the clinical terminology as the
people suffering from migraines) have their own prediction flags such
as aura (perceptual disturbance experienced by the patient before the
pain) or prodromic symptoms (subjective and unspecific perceptual
disturbances). However, these symptoms can appear at any time from 48
to 6 hours before the onset of the migraine, discarding these as good
predictors. Our hypothesis is the following: if we understand the
changes that happen to the hemodynamic variables, we could predict the
onset of a migraine. If the prediction time is long enough to reach
the times of the pharmacokinetics of the drugs, we could anticipate
the intake of these to abort the migraine.

\ignore{Knowing the changes that happen in the hemodynamic
variables, the prediction of the migraine disease is possible, and
reaching sufficiently long prediction horizons in the pharmacokinetics
would anticipate the intake of drugs to abort the migraine.}

The aim of this paper is to propose a methodology to show the limits
of prediction of symptomatic crises using state-space models. The main
goal of this work is not in the prediction itself, that has been
already proved by the authors, but in the mechanisms applied to
increase the accuracy of predictive models by tuning their
parameters. In this paper, we also show how predictions can be
improved by removing spurious and noisy data in the input data
set. Predictive algorithms frequently applied in the literature to
static datasets~\cite{kutur2012improved,huang2013online}, where there
is no data loss and signals are less noisy. This study will use real
data gathered from a WBSN, that imposes severe constraints in the
processing of noisy and unreliable data. Thus, the proposed
methodology will study the best options for prediction according to
the availability or status of sensors and the desired horizon using
data from a real ambulatory study.

The remainder of this paper is as follows. \secref{methods} explains
the methodology followed to gather the data and its management, as
well as the description of the parameters used in the algorithms
envisioned to solve our problem. \secref{results} shows the results
obtained and their discussion. Finally, some conclusions of this work
are drawn in~\secref{conclusions}.


\section{Methods}
\label{sec:methods}
\begin{figure}
 \centering
 \includegraphics[width=1.0\columnwidth]{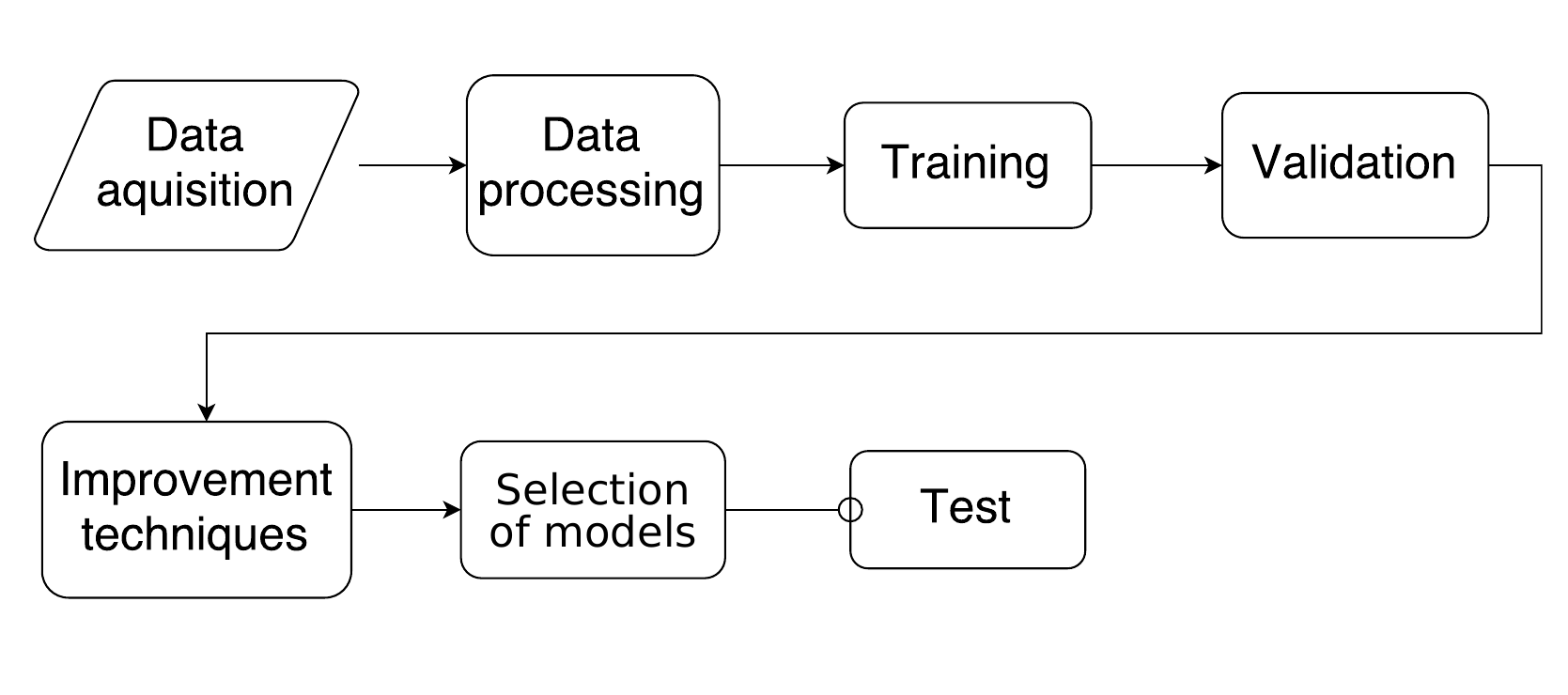}
 \caption{General overview of the proposed methodology.}
 \label{fig:dia_metho}
\end{figure}

This work presents a methodology to improve prediction models for
chronic diseases with symptomatic crises. It also analyzes the
prediction limits when applied to a real case study of a chronic
disease with symptomatic crises, the migraine. Due to the complexity
of an ambulatory study like the one presented in this paper
(recruitment of patients, deployment of a large number of monitoring
devices, long monitorization time, etc.), this work is focused only on
the migraine disease. In the opinion of the authors, the methodology
presented in this work is fully applicable to other chronic diseases
drawing subjective pain symptomatic curves. The diagram
in~\figref{dia_metho} presents an overview of the steps of the
proposed methodology.

Firstly, \subsec{data} relates the main characteristics of the data
used and the developed processing techniques. Next, training and
validation of the models are shown in \subsec{models}. Finally, a very
critical block of our methodology is presented: the development of
improvement techniques, shown in \subsec{improving_meth}. These
techniques will be applied after obtaining the predictive model, in
order to increase the accuracy and the horizon length. Results
obtained by our techniques support the importance of deriving a
methodology like the one proposed in this paper.

\subsection{Data}
\label{subsec:data}

Four hemodynamic variables have been monitorized in migraineurs during
24 hours per day: heart rate (HR), electrodermal activity (EDA), skin
temperature (TEMP) and peripheral capillary oxygen saturation
(SpO2). A multivariable analysis of these signals is applied to
predict a migraine crisis. In addition to the hemodynamic variables,
the subjective pain has been manually registered by patients to
correlate the real pain with the biometric signals and to train the
predictive models.

Changes in these hemodynamic variables regulated by ANS are related in
the clinic literature to the migraine. For instance,
Hassinger~\etal~relate the cardiovascular response to the
migraine~\cite{hassinger1999cardiovascular} and Vollono~\etal~do the
same with the heart rate variability during the
sleep~\cite{vollono2013heart}. Kewman~\etal, for example link changes
in the skin temperature with migraines, as other authors
do~\cite{kewman1980skin}. In a previous study, these variables have
demonstrated to be good predictors of the
migraine~\cite{pagan2015robust}. Passchier shows also changes in the
electrodermal activity in migraine sufferers
in~\cite{passchier1993abnormal}. Regarding the SpO2, Lovati shows
in~\cite{lovati2012breathing} how blood oxygenation during sleep was
significantly higher among headache patients with respect to controls.

\begin{table}
\caption{Data acquisition parameters.}
  \centering
  \resizebox{\columnwidth}{!}{ 
    \begin{tabular}{ccccc}
      & Placement & Sampling (Hz) & Precision & Data-24h (KB) \\ 
      TEMP & Armpit & 1 & 0.0223 \degree C & 126.6 \\ 
      EDA & Arm & 1 & 0.0062 $\mu$ S & 126.6 \\
      ECG (HR) & Breast & 250 (0.1) & 4 ms (1 bpm) & 31640.6 (12.7) \\ 
      SpO2 & Finger & 3 & 1 \% & 253.1 \\ \midrule
      \multicolumn{4}{c}{Total (MB)} & 31.4 (0.51) \\ 
     \end{tabular}
  }
  \label{tbl:datachar}
\end{table}

Once the patients have signed the informed consent (the protocol for
the clinical study that was approved by the Local Ethics Committee of the
hospital), the monitorization phase begins. Two sensing motes are
used: i) PLUX-Wireless Biosignals
\cite{BIOPLUXws} to acquire EDA, skin
temperature and ECG signals, and ii) Nonin Onyx II \cite{NONINws} to
acquire SpO2. \tblref{datachar} summarizes the placement of sensors,
their data acquisition rate, accuracy, and the amount of data gathered
during 24 hours of monitorization. Despite the HR is used for
modeling, this is calculated offline from the ECG signal; this fact
reduces the amount of data to process from 31.4 MB to 0.51 MB per day.

Patients indicate through an electronic form in an Android smartphone
the beginning and the end of the symptomatic crisis. They also mark
the relative changes in pain intensity or punctual pain levels during
the migraine crisis (several marks during the migraine).  These
relative changes are not limited in a numbered
scale~\cite{pagan2015robust} (from $-2^{32}+1$ to $2^{32}-1$). In
addition, patients mark a global pain that defines the whole migraine,
this time in a normalized and limited scale
0--10~\cite{kellogg2012association}, in order to verify that the
crisis corresponds to a migraine or another kind of headache. The two
sensing motes send the data to the smartphone via Bluetooth and then
the data are transmitted to a Cloud storage system. Data processing as
well as optimization and predictive algorithms run on a remote PC or
server.

\begin{figure}
 \centering
 \includegraphics[width=1.0\columnwidth]{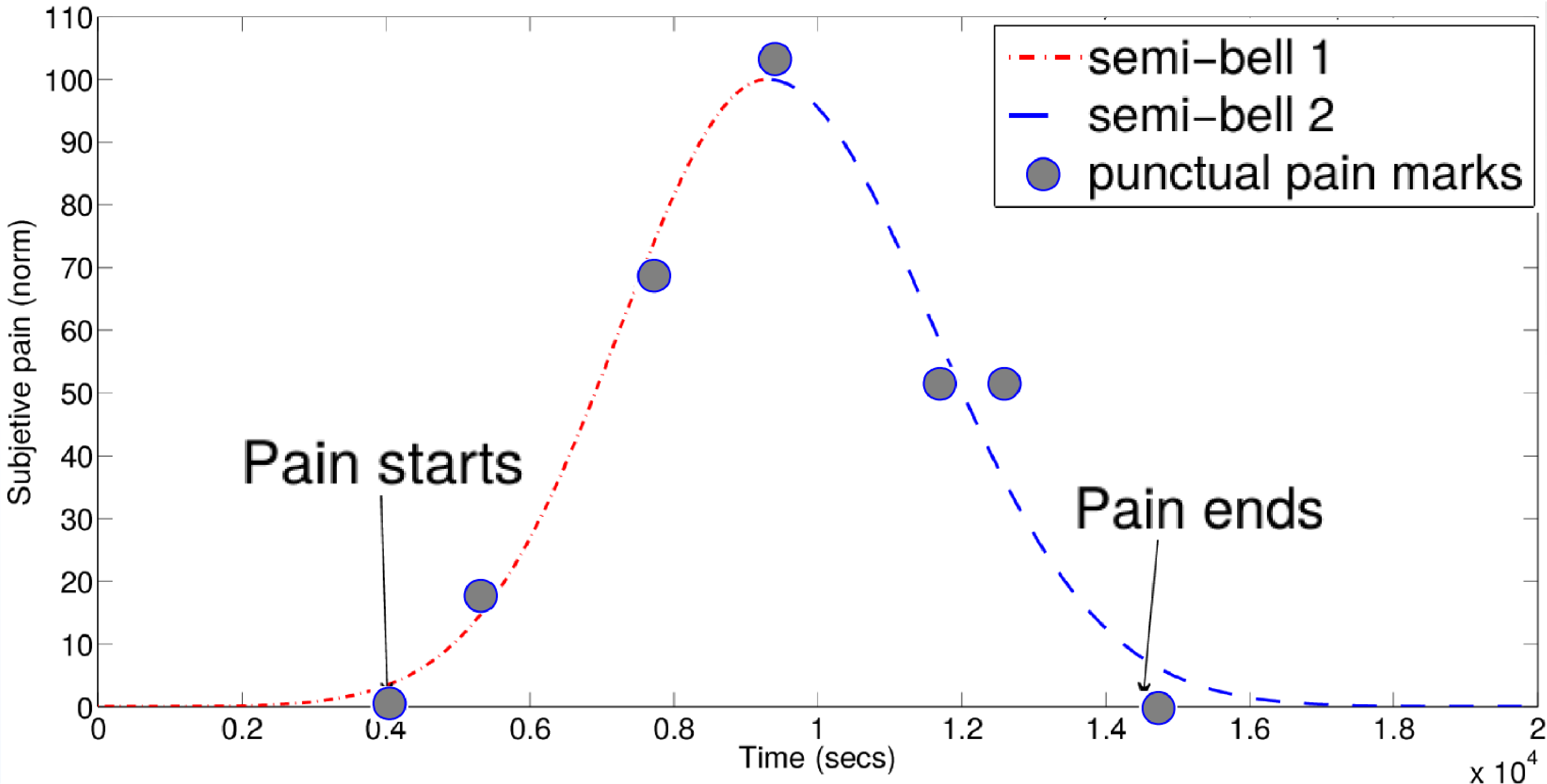}
 \caption{Modeling of subjective pain evolution curve.}
 \label{fig:gaussianPain}
\end{figure}

The punctual relative pain levels indicate the subjective pain
intensity. The maximum represents the highest pain, and it will be
different for each migraine and patient. Patients do not know if their
current pain is the maximum or not. Hence, the use of an unlimited
scale allows marking high values to prevent saturation. Each curve is
normalized (0 to 100\%) and modeled as two semi-Gaussian curves. These
curves have shown a good fit to the points marked by the patients. The
parameters necessary to define such symptomatic curve
are \{$(\mu_{1}, \sigma_{1}),
(\mu_{2}, \sigma_{2})$\}~\cite{pagan2015robust}. The symptomatic curve
includes the aura (if it exists) because it reflects some changes in
the migraine process and this is considered a symptomatic process. An
example is shown in \figref{gaussianPain}.

Signals are synchronized before running the algorithms. After that, in
order to recover disruptions in data, a Gaussian Process Machine
Learning (GPML) procedure is followed. This process is based on the
work developed by Rasmussen~\cite{Rasmussen06gaussianprocesses} and
their tool~\cite{GPMLtool}. Signals are synchronized, and the
disruptions in the data are repaired using a Gaussian likelihood
function by GPML. The normalized root mean square error (NRMSE), also
called fit, is the metric used to calculate the goodness of the
fitting of GPML with the available data (some data are lost during
transmission because sensors can disconnect or the wireless
transmission link is noisy). The fit is defined as:

\begin{equation} 
 \begin{aligned}
 fit = 100 \times \left(1-\frac{\|y-\hat{y}\|}{\|y-mean(y)\|}\right)
 \label{eq:metric}
 \end{aligned}
\end{equation}

where $y$ is the real (Gaussian modeled) symptomatic curve, and
$\hat{y}$ is the predicted one.

\begin{figure*}
\captionsetup[subfigure]{labelformat=empty}
  \hspace{-1.5cm}
 \subfloat[]{
  \includegraphics[width=0.46\textwidth]{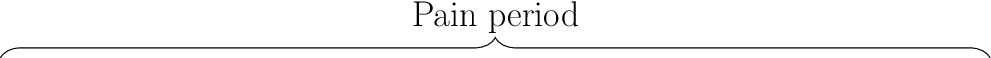} 
  }
  \vspace{-0.8cm}
 \centering
 \captionsetup[subfigure]{labelformat=empty}
 \subfloat[]{
  \includegraphics[width=\textwidth]{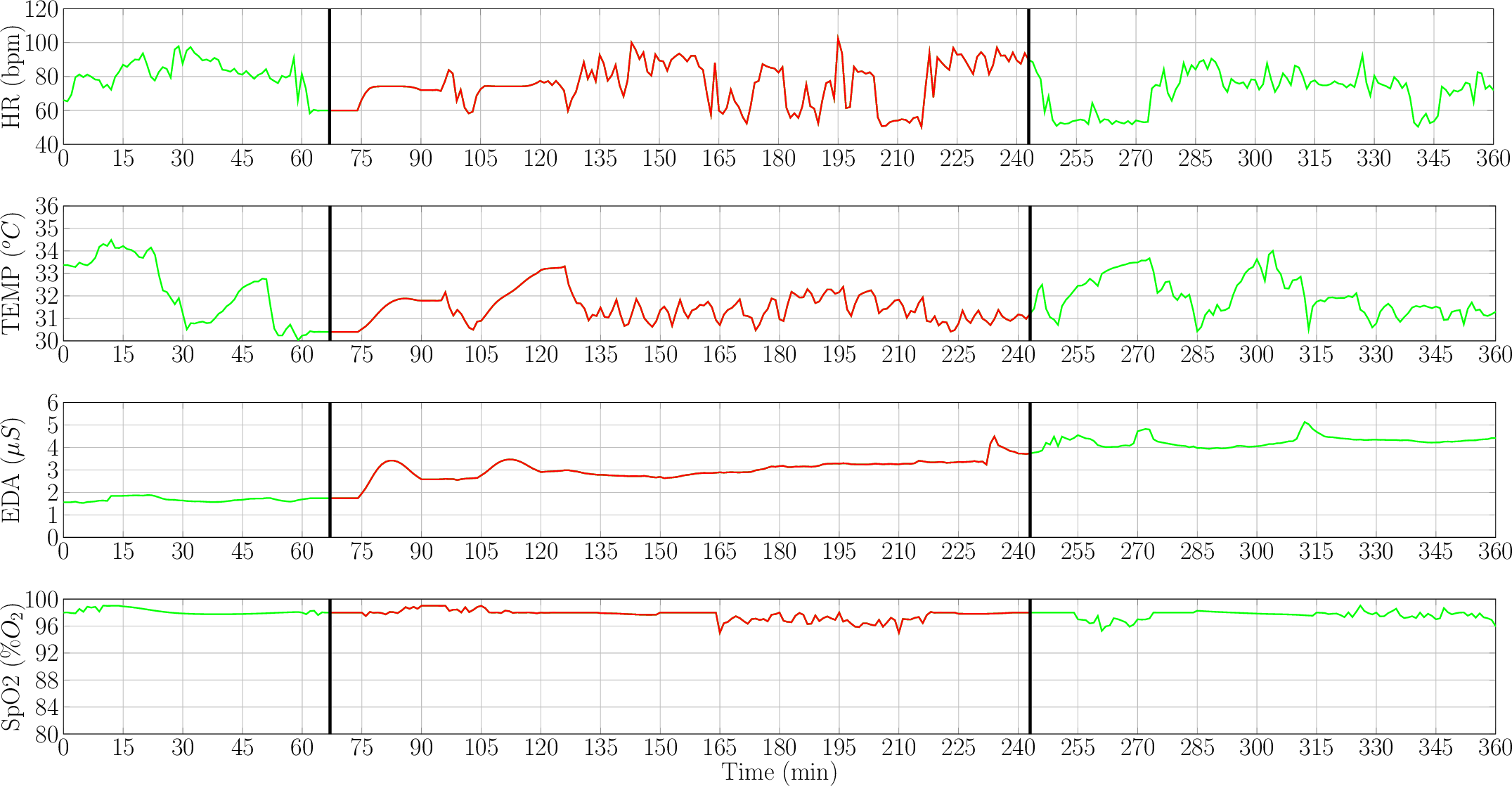} 
 }
\vspace{-0.5cm}
 \caption{Hemodynamic variables after synchronization and
 preprocessing during a migraine episode (red curve between vertical bars).}
 \label{fig:gpml}
\end{figure*}

After the synchronization, the time between samples is set to 1 minute
for all signals. \figref{gpml} shows the four hemodynamic signals
during an asymptomatic period (green lines) and a migraine event (red
lines between vertical bars) in the middle. These data have been
synchronized and repaired using the GPML. The average fit achieved for
all the signals in \figref{gpml} using the GPML is $81.9\%$.

For this paper, in order to show results for the methodology proposed
in \subsec{improving_meth}, data from two patients have been selected
from the monitoring database (labeled as Patient A and B). Data from
Patient A correspond to a young female patient that suffers from
migraines with aura and does not undergo medical treatment. 20
migraines have been acquired in two different experimental periods
(nearly one month each). Data from Patient B correspond to a middle
aged female patient that suffers from migraines without aura and
undergoes preventive medical treatment. 12 migraines have been
acquired in one experimental period (almost a month). The training
dataset for Patient A and B was of M=15 and M=8 randomly chosen
migraine events, respectively.

\subsection{Models}
\label{subsec:models}

It is well known that migraines are a sequence of neurological stages:
i) prodromic symptoms, ii) aura phase, iii) the pain itself and iv)
finally a postdromic stage~\cite{burstein2015migraine}. As
aforementioned, the intake time of the drugs used to stop the symptoms
of the migraine is of critical importance. The earlier the intake, the
more the effectiveness, because when the pain cycle begins, there is
an activation of the trigeminal nucleus and it is much more difficult
to stop it~\cite{goadsby2008early}. Thus, the success of the medicine
to stop the pain strongly depends on the prediction horizon; hence, a
methodology to achieve the maximum prediction horizon is needed.

The N4SID state-space algorithm~\cite{van1994n4sid} has been chosen
for its accuracy in modeling other biomedical processes, such as
Cescon presented in~\cite{4145031} or Facchinetti
in~\cite{facchinetti2011detecting}, both studies about diabetes. N4SID
models have been used in other previous works in the literature of
bioinformatics applications, reaching good results. For instance, to
estimate infections in populations, like Tan~\etal~did
in~\cite{tan2000estimation} for HIV, or Hooker~\etal~did
in~\cite{hooker2010parameterizing} for infectious diseases. The N4SID
algorithm has shown also good results in the biomedical area as
detector of anomalies in the electrocardiogram signal, as
Munevar~\etal~demonstrate in~\cite{munevar1999detection}.

In this work, the N4SID algorithm has been computed using the System
Identification Toolbox of the MATLAB software~\cite{matlab}.

\subsubsection{Training the models}
\label{subsubsec:training}

A state-space model is a mathematical representation that describes an
output (or multiple outputs) as the relation of a set of inputs and
state variables by difference equations. These states are
immeasurable. The current and future outputs are related, through the
system, with past and current inputs. The N4SID is a stochastic model,
represented in the general form as a multi-input multi-output (MIMO)
linear time-invariant system (LTI)~\cite{hangos2001intelligent} as
in \eref{n4sid}:

\begin{equation}
 \begin{aligned}
 x_{k+1} = Ax_{k}+Bu_{k}+w_{k} \\
 y_{k} = Cx_{k}+Du_{k}+v_{k}
 \label{eqn:n4sid}
 \end{aligned}
\end{equation} 

In our case, $u_k$ are 4 hemodynamic inputs and $y_k$ is 1 output
(pain level), both at time step $k$. $A$ is the state transition
matrix and relates the next state ($x_{k+1}$) to the current one
($x_{k}$); $B$ relates the next state to the current inputs
($u_{k}$); $C$ relates the current state to the current output
($y_{k}$), and $D$ equals zero in our case. $v_k$ and $w_k$ are
immeasurable white noises.

The metric used to evaluate the accuracy of the models is the
aforementioned fit. In the training process, the N4SID order $nx$
(size of the square matrix $A$) and the number of samples from the
past inputs and the output are chosen by the algorithm as the ones
that achieve the best fit. To do this, the process runs in a triple
loop looking for the best parameters. The inner loop chooses the order
of the models. The one in the middle chooses the backward window to
get information from past inputs. The outer loop only fixes the
prediction horizon as shown in this pseudo-code:

\lstset{float,language=Matlab,basicstyle=\scriptsize}        
\begin{lstlisting}[frame=single]          % Start your code-block

prevFit = -Inf; %Previous fit achieved

% For six future, sixty past horizons and ten orders 
for futWin = 10:10:60
  for pastWin = 0:5:120
     opts = n4sidOptions(futWin, pastWin);
     for nx=1:10         
        % Calculate the model
        stateSystem = n4sid(data, nx, opts);

        % Calculate the fit
        fit = compare(data,stateSystem,futWin);

        if fit > prevFit
           % Paremeters of the current best model 
           prevFit = fit;        
           bestPast(futWin) = pastWin;
           bestOrder(futWin) = nx;
        end
     end
  end
end
\end{lstlisting}

In addition, a parallel study for feature selection has been
performed. Models have not only been trained with four hemodynamic
inputs, but also with the combinations in triads of them; in total, we
checked 5 sets of features. From these experiments, we will obtain the
features that better describe a migraine per patient.

After training the models, 240 combinations are checked to select the
best one per future horizon and per migraine event, and per set of
features.

\subsubsection{Validation of models}
\label{subsubsec:validation}

In the validation process, we look for the best models to predict
migraines using the cross-validation criteria: each model $M_{i}$,
$i=1,2,\ldots,M$, obtained from the $i$-th migraine is validated
against the other $j$-th migraines, with $i \neq j$. The validations
are performed for the same horizon for which the model was trained. We
compare two models $M_{i}$ and $M_{j}$ regarding their average
fit. The average fit of each model is calculated from the $M-1$
validation. The better the average fit in validation, the better the
model.

\begin{figure}
 \centering
 \includegraphics[width=1.0\columnwidth]{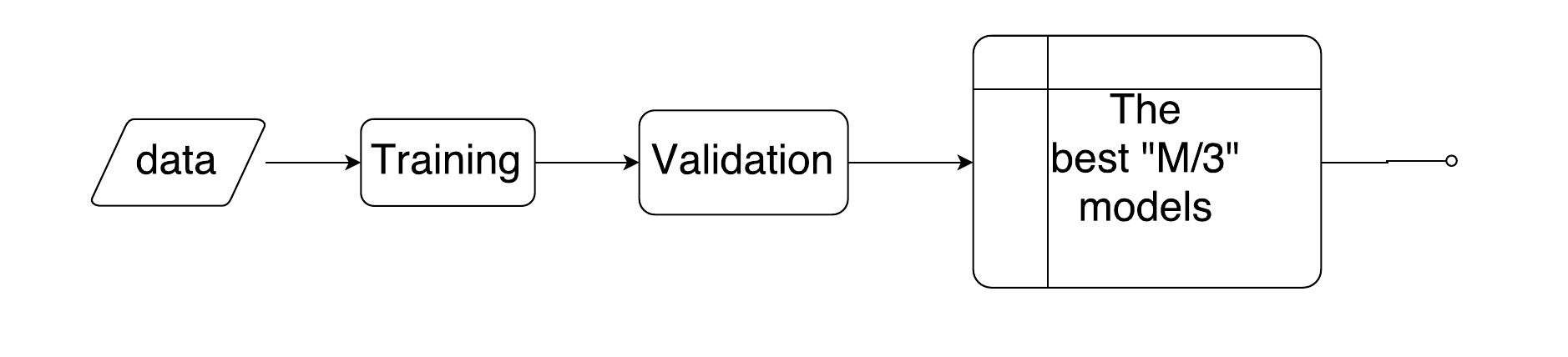}
 \caption{Basic scheme to select the best models for each patient.}
 \label{fig:dia_val_base}
\end{figure}

A model is considered good when it is able to validate at least
$\lceil M/3\rceil$ of the migraines from the dataset at a given
fit. More than one model is selected in order to calculate an average
prediction and to avoid any bias. The criterion followed is to select
the best $\lceil M/3\rceil$ models trained. \figref{dia_val_base}
represents the basic scheme of training and validation. The last
module in the figure represents the model selection that implements
the proposed methodology.

\subsection{Improving results}
\label{subsec:improving_meth}

\begin{figure}
 \centering
 \includegraphics[width=1.0\columnwidth]{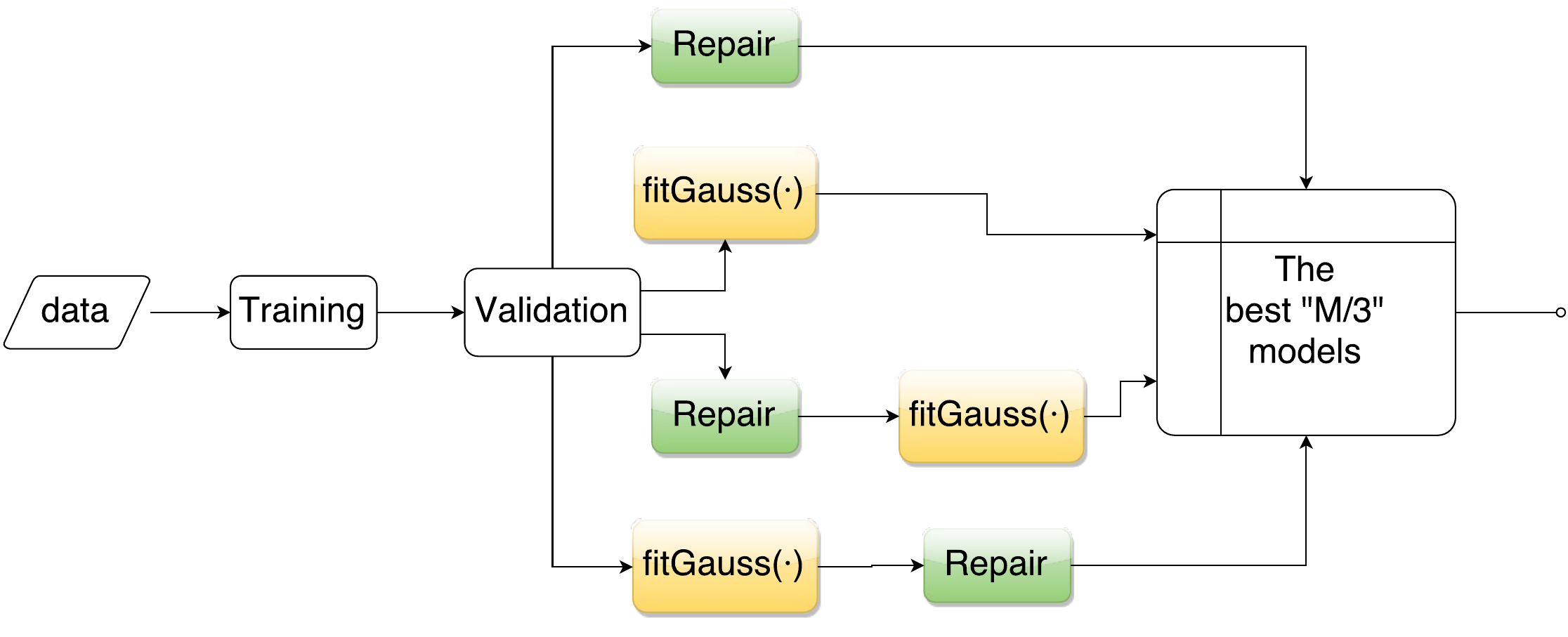}
 \caption{Schemes proposed to be used in the methodology.}
 \label{fig:dia_val_all}
\end{figure}

This block, from the scheme shown in \figref{dia_metho}, implements a
sequence of processes to improve the prediction. The predictions
obtained by the N4SID models have difficulties maintaining a
constant value, and they tend to oscillate around the zero value when
no symptomatic crisis is detected. This fluctuation causes an
artificial reduction in the fit. The blue curve
in \figref{repairProccess_pac11_M1r4_m24} represents a prediction with
fluctuations (the original symptomatic curve is the black one). These
oscillations can be easily detected and removed. To do this, two
methods are evaluated: i) reparation of the prediction, and ii)
Gaussian fitting as the original symptomatic crisis was modeled. These
methods are applied to the basic scheme of \figref{dia_val_base}. All
the possibilities studied are shown in \figref{dia_val_all}. Each one
of the four branches represents a scheme to improve the predictions.

\subsubsection{Reparation of the prediction}
\begin{figure*}
 \subfloat[]{
  \includegraphics[width=0.33\textwidth]{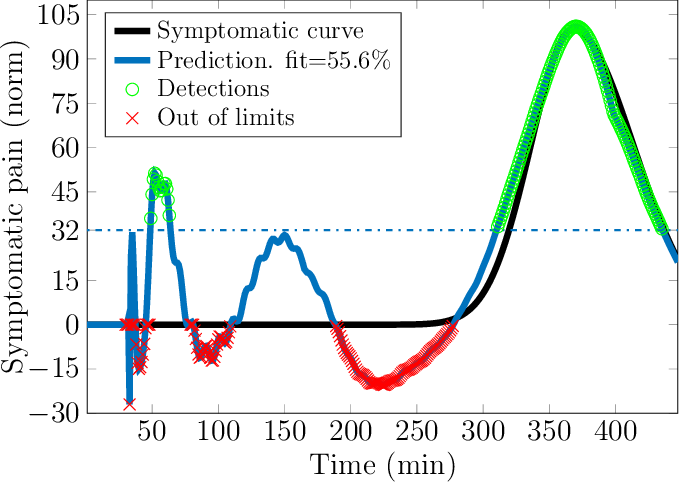} 
  \label{fig:repairProccess_pac11_M1r4_m24}
 }
 \subfloat[]{
  \includegraphics[width=0.33\textwidth]{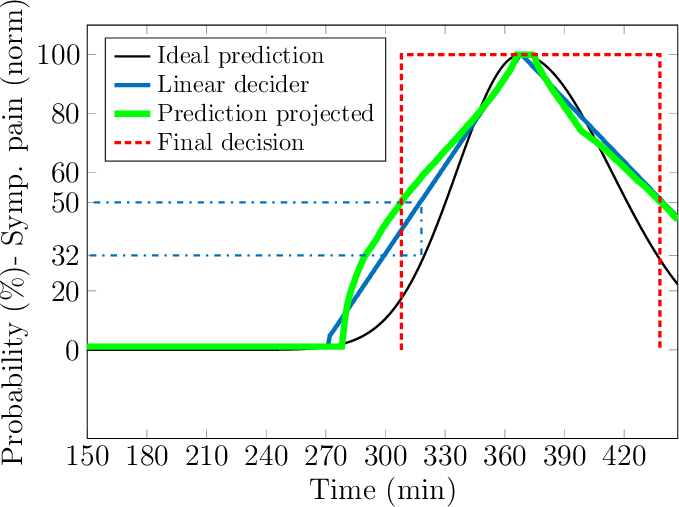} 
  \label{fig:decider_pac11_M1r4_m24}
 }
 \subfloat[]{
  \includegraphics[width=0.33\textwidth]{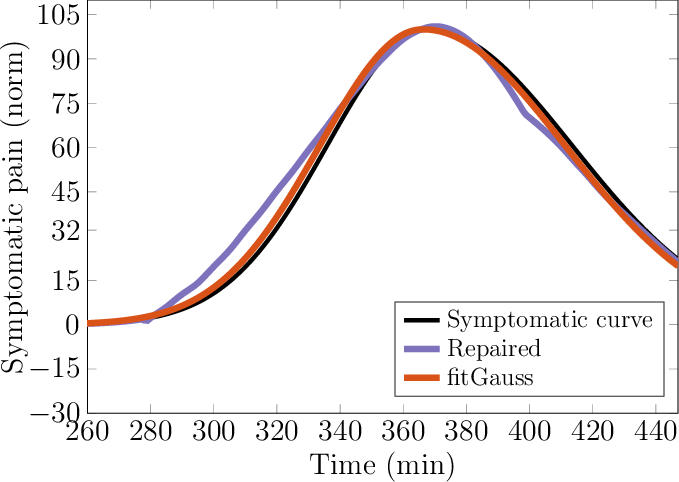} 
  \label{fig:repairedAndFitG_pac11_M1r4_m24}
 }
 \caption{Improving the predictions. ({\bf a}) Prediction over real
 symptomatic curve. Detection of events, removal of oscillations and
 false positive detected; ({\bf b}) Probability curve of pain
 occurrence from the repaired prediction over the ideal probability
 curve. Final detection time limits; ({\bf c}) Result after repairing
 the prediction and fitting the prediction to two semi-Gaussian curves.}
\end{figure*}

In order to illustrate these
processes, \figsref{repairProccess_pac11_M1r4_m24}{decider_pac11_M1r4_m24}
show how to repair a prediction. False positives are detected using a
level and a time threshold. Firstly, those values out of limits (below
zero and above the maximum) are marked with red \emph{x}, as shown
in \figref{repairProccess_pac11_M1r4_m24}. Then, negative values are
set to zero, and the rest of outliers are set to the maximum. After
that, the level threshold is applied. This process marks as detections
those values above 50\% of the probability of occurrence (green
circles in \figref{repairProccess_pac11_M1r4_m24}), using the linear
decider explained below. The 50\% of pain probability is projected to
a level of 32 over the ideal prediction (same as the original
symptomatic curve). The blue dotted line represents this
in \figref{decider_pac11_M1r4_m24} and extends
through \figref{repairProccess_pac11_M1r4_m24}.

Finally, the time threshold is applied. If the distance between the
farthest points is lower than 60 minutes (enough to detect if a
migraine attack occurs or not), it is considered as a false
positive. These points are
removed. In \figref{repairProccess_pac11_M1r4_m24} the left detection
is removed.

As a result, the repaired prediction is represented
in \figref{repairedAndFitG_pac11_M1r4_m24} (purple curve). It is worth
noting that the fluctuation that appears in the middle of the curve
was not detected by the threshold level.

In the following, we present how the linear decider works. The
use of this decider was initially introduced in a previous
work~\cite{pagan2015robust}, and here we summarize the main
characteristics to help the reader to understand the next stages of
this research.

The linear decider will detect a migraine event when the probability
of occurrence of a detection exceeds the 50\% of probability. This
linear decider (blue triangle in \figref{decider_pac11_M1r4_m24})
ranges from 0\% (minimum pain intensity in the normalized symptomatic
Gaussian curve) to 100\% (maximum pain intensity in the normalized
symptomatic Gaussian curve) of probability
(see~\subsec{data}). Therefore, the linear decider projects the
repaired prediction (blue signal
in \figref{repairProccess_pac11_M1r4_m24}) to a probability of
occurrence curve (green curve in \figref{decider_pac11_M1r4_m24}). The
linear decider uses a linear function as the projection function. As a
result, the migraine detected (all those values higher than the 50\%
of probability of occurrence) is bounded by the red dotted line
in \figref{decider_pac11_M1r4_m24}.

\subsubsection{Gaussian fitting}
\figref{repairedAndFitG_pac11_M1r4_m24} also illustrates the
result of applying the Gaussian fit (orange curve). This process fits
the prediction to two semi-Gaussian curves, with reference at the
maximum of the prediction. With the aim of finding the original bells,
the prediction is first normalized and then fitted.

The impact of the combination of both processes (repair and Gaussian
fitting, depicted in the two lower branches of \figref{dia_val_all}),
is also analyzed in~\secref{results}.

\subsection{Fitting and prediction criteria}
\label{subsec:criteria}
The goodness of the fit and the prediction horizon can be used as
criteria to select the models. Selecting one criterion automatically
sets the other. Setting the fit, we can follow a more conservative
approach that reduces the prediction horizon but improves the
confidence in the models. However, setting the prediction horizon can
achieve farthest predictions by loosing accuracy. In the following
section, we choose first a minimum fit as requirement for the model
selection. Later, a strategy is shown to select the models according
to the prediction horizon.

A fit of 70\% has been selected as the threshold of similarity to
consider a model as good candidate.


\section{Results}
\label{sec:results}

In this section we present the experimental results obtained by our
proposed methodology, designed to enlarge the prediction horizon of
predictive models for symptomatic crises. As stated before, this
method has been tested with N4SID, a well-known state-space based
algorithm. Firstly, we present the results of model
training. Secondly, we show the validations of these models, where all
the improvement schemes presented in \figref{dia_val_all} are
studied. Finally, one scheme is selected and the models are tested
with new signals (migraines not used in the training and validation
sets).

Along this section, our criterion has selected a 70\% for the fit
value. This is a conservative setup that will improve the confidence
in the models (see \subsec{criteria}). The alternative approach (to
set the prediction horizon by loosing accuracy) is also tested
in \subsec{strategy}.

\subsection{Training the models}
\label{subsec:training}

\begin{figure*}
 \centering
 \subfloat[]{
  \includegraphics[width=0.5\textwidth]{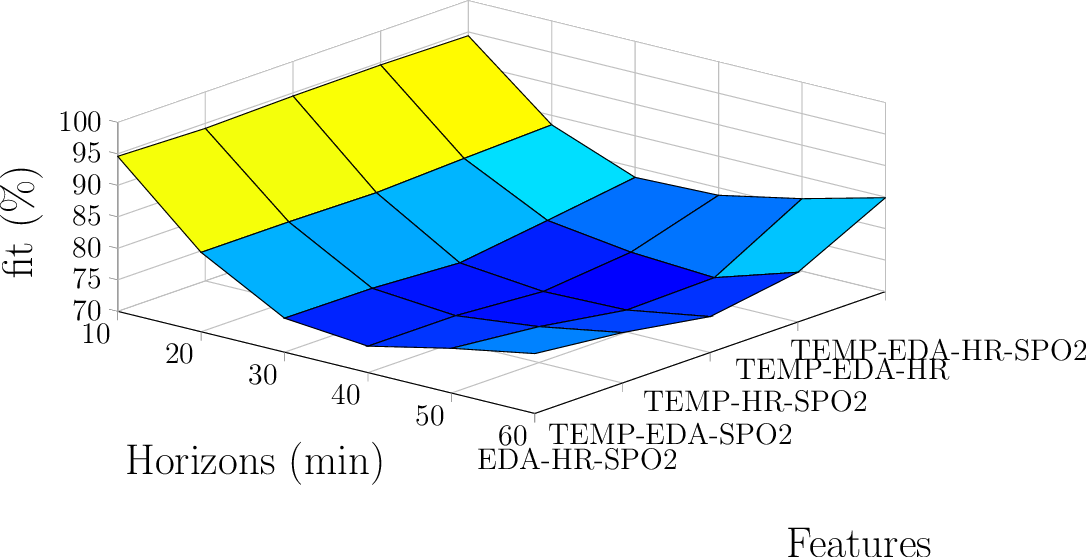} 
  \label{fig:pac2_training_3D}
 }
 \subfloat[]{
  \includegraphics[width=0.5\textwidth]{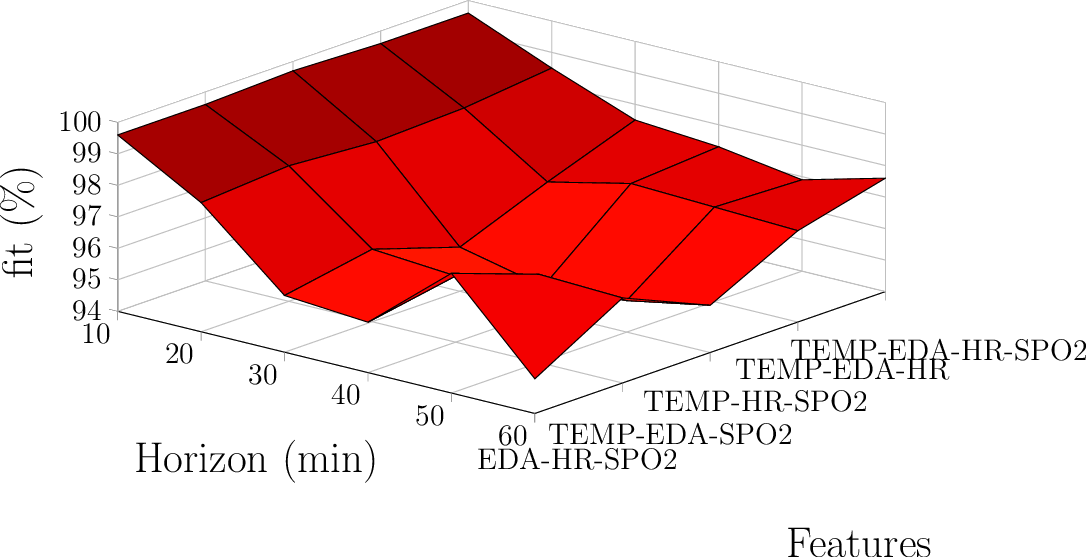} 
  \label{fig:pac11_training_3D}
 }
 \\
 \vspace{-0.25cm}        
 \captionsetup[subfigure]{labelformat=empty}
 \subfloat[]{
  \includegraphics[width=\textwidth]{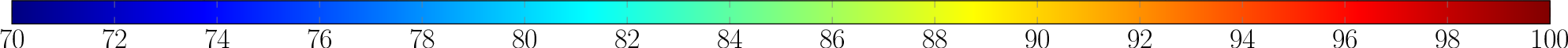} 
  \label{fig:colorbar}
 }
 \caption{Average fits for training. Dependence with the future
 horizon and selected variables. ({\bf a}) Data from Patient A; ({\bf
 b}) Data from Patient B}
\end{figure*}

As mentioned in \subsubsec{training}, each migraine has been trained
for 6 different horizons and 5 different feature
combinations. \figsref{pac2_training_3D}{pac11_training_3D} summarize
the training results for patients A and B respectively. Each value on
the surface of these graphs represents the average of fits over all
the trained models (M = 15 models for Patient A, and M~=~8 for Patient
B). For Patient A, the fit decreases more quickly than for Patient
B. In addition, the training results for patient B are almost 15-20
points higher than the results for Patient A. This can be explained by
the higher amount of data lost during monitorization of Patient A (in
spite of the usage of GPML).

\begin{table*}
\caption{Training results for the TEMP-EDA-HR-SpO2 features set and 40
minutes forward horizon for patients A and B.}
  \centering
  \resizebox{\textwidth}{!}{ 
\begin{tabular}{ccccccccccccccccccccccccc}
\toprule
& \multicolumn{15}{c}{Patient A} & & \multicolumn{8}{c}{Patient B}\\ \midrule
& $M_{1}$ & $M_{2}$ & $M_{3}$
& $M_{4}$ & $M_{5}$ & $M_{6}$
& $M_{7}$ & $M_{8}$ & $M_{9}$
& $M_{10}$ & $M_{11}$ & $M_{12}$
& $M_{13}$ & $M_{14}$ & $M_{15}$ &
& $M_{1}$ & $M_{2}$ & $M_{3}$
& $M_{4}$ & $M_{5}$ & $M_{6}$
& $M_{7}$ & $M_{8}$\\ \midrule
$fit (\%)$ & 84,4 & 84,1 & 88,1 & 75,6 & 70,6 & 85,0 & 86,7 & {\bf
65,8} & 78,9 & 82,6 & 79,2 & {\bf 67,4} & 80,8 & 81,0 & 72,0 &  & 97,5
& 99,6 & 95,9 & 99,5 & 90,0 & 97,0 & 99,9 & 98,1 \\ 
$ph$ (min) & 25 & 105 & 60 & 40 & 30 & 30 & 70 & 75 & 15 & 100 & 60 &
95 & 20 & 105 & 90 &  & 12 & 30 & 14 & 20 & 25 & 25 & 15 & 18 \\  
$nx$ & 6 & 4 & 7 & 8 & 5 & 9 & 5 & 6 & 10 & 9 & 7 & 8 & 7 & 4 & 7 &  &
6 & 9 & 7 & 6 & 10 & 8 & 10 & 9 \\ \bottomrule 
\end{tabular}
}
\label{tbl:training_40_min_all_pac}
\end{table*}


In \figsref{pac2_training_3D}{pac11_training_3D}, maximum fits are
reached for lowest horizons (10 and 20 minutes). The fit decreases
with the horizon, but also depends on the features selected. The
highest values of fit are reached for the combination of the four
available biometric variables. It is worth to mention that a valley is
found around the prediction horizon of 40 minutes. Surprisingly, this
occurs for the TEMP-HR-SpO2 feature combination in both patients. As
the number of individuals is not enough, this should not be considered
as a conclusion. At this point, the fit for Patient A is $73.2\%$, and
$94.8\%$ for Patient B. This suggests that the time window for
prediction is larger for Patient B than for Patient A. Additionally,
fits increase with prediction horizons larger than 40 minutes (50 and
60 minutes); this is due to overfitting during training
(see \subsec{validation}). It seems that our modeling approach in the
training stage reaches the limit for the migraine prediction at 40
minutes.

\tblref{training_40_min_all_pac} shows the training results
for each model for patients A and B. This table summarizes the fit
reached by the models, the past horizon ($ph$, in minutes) required to
train them, and the order of the matrices required to reach the best
fit. For the sake of space, only results for the horizon of 40 minutes
and for the TEMP-EDA-HR-SpO2 combination of features are shown. This
setup corresponds to the minimum training error. The fits are $78.8\%$
and $97.2\%$ in average for patients A and B, respectively.

As can be seen, fits reached are high for both patients, and they
are always over 70\% (except for two Patient A cases marked in
bold in \tblref{training_40_min_all_pac}). However, models require
large matrices (larger than order $nx=7$) in most of the
cases. Despite the high orders, past horizons are low for Patient B
(they are always lower than 30, 20 minutes in average); but they are
high for Patient A (61 minutes in average and up to 105 minutes
backward). For the remaining future horizons and feature sets, the
average fit in training keeps high, always over the 70\% for both
patients. No correlation has been found between the order of matrices
and the number of past inputs.

In (\subsec{validation}) we present the results for model
validation. Here, trained models are tested as predictors of the
other symptomatic crises of the training dataset. In the following
section, we will also analyze the overfitting effect.

\subsection{Validations of models}
\label{subsec:validation}

In this section we show the results of performing cross-validation between
models, as mentioned in \subsubsec{validation}. The main objective of
this section is to discard overfitted models. In this way, we will
find those models that reach the longest prediction horizon. Results
have been obtained for the 6 different prediction horizons and the 5
feature combinations.

\tblref{num_models_average_all_horizons} represents the number of useful
models with average fit over all the cross-validations exceeding
70\%. As the average prediction is calculated over more than one
model, this analysis will help on the selection of the models. It is
considered that, at least, one third of the models must validate with
high average fit to choose a feature set as relevant (for each
prediction horizon). According to the results
in \tblref{num_models_average_all_horizons}, no difference appears
between the selected features for a forward horizon of 10 minutes. In
general, no model is able to validate for higher horizons than 20 and
30 minutes for patient A and B respectively. This confirms that the
valley in training in \figsref{pac2_training_3D}{pac11_training_3D}
marks the limit of prediction for state-space models, and models
trained over 40 minutes are overfitted.

The four-features combination is always the worst combination. For
Patient A and 20 minutes forward horizon, the combinations of three
features, except for TEMP-HR-SpO2, show 5 available models (in the
limit of our criterion to consider the features as relevant). For
Patient B, all the combinations of features look good (more than 3
models over the average fit of 70\% in this case) for 20 minutes, but
only the TEMP-HR-SpO2 feature combination is useful for 30 minutes.

\begin{table}[htbp]
\caption{\# Useful models after validation.}
        \centering
         \resizebox{\columnwidth}{!}{
         \begin{tabular}{cccccccccccccc}
         \toprule
         & \multicolumn{6}{c}{Patient A} & & \multicolumn{6}{c}{Patient B} \\ \midrule
         Features / Horizon (min) & 10 &  20 & 30 & 40 & 50 & 60 & & 10 &  20 & 30 & 40 & 50 & 60 \\ \midrule 
         TEMP-EDA-HR-SpO2 & 15 & 4 & 0 & 0 & 0 & 0 &  & 8 & 4 & 1 & 0 & 0 & 0 \\ 
         TEMP-EDA-HR & 15 & 5 & 0 & 0 & 0 & 0 & &  8 & 6 & 1 & 0 & 0 & 0 \\ 
         TEMP-EDA-SpO2 & 15 & 5 & 0 & 0 & 0 & 0 &  & 8 & 6 & 2 & 0 & 0 & 0 \\
         TEMP-HR-SpO2 & 15 & 7 & 0 & 0 & 0 & 0 &  & 8 & 7 & 4 & 0 & 0 & 0 \\ 
         EDA-HR-SpO2 & 15 & 5 & 0 & 0 & 0 & 0 & & 8 & 7 & 1 & 0 & 0 & 0 \\ 
         \bottomrule  
         \end{tabular}
         }
\label{tbl:num_models_average_all_horizons}
\end{table}

As aforementioned in \subsec{training}, high fits in training do not
assure good models, and some of them must be discarded in the
validation phase.

\begin{figure*}
 \centering
 \subfloat[]{
  \includegraphics[width=0.33\textwidth]{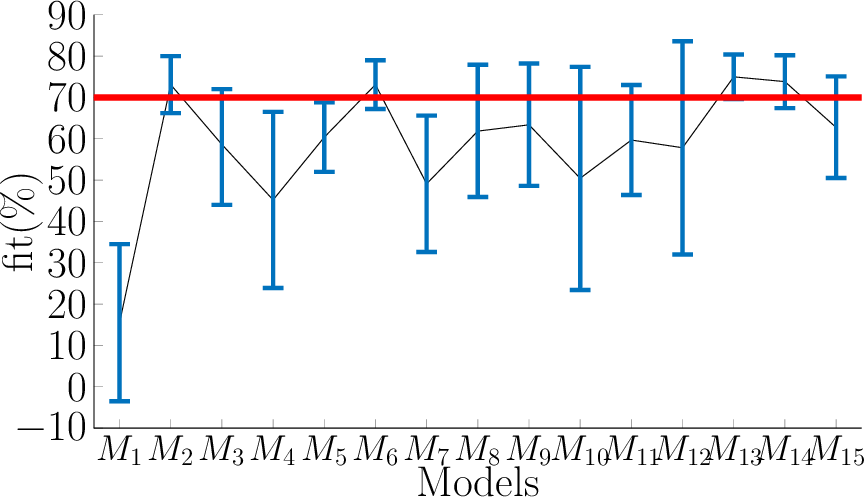} 
  \label{fig:pac2_val_20_min_4feat}
 }
 \subfloat[]{
  \includegraphics[width=0.33\textwidth]{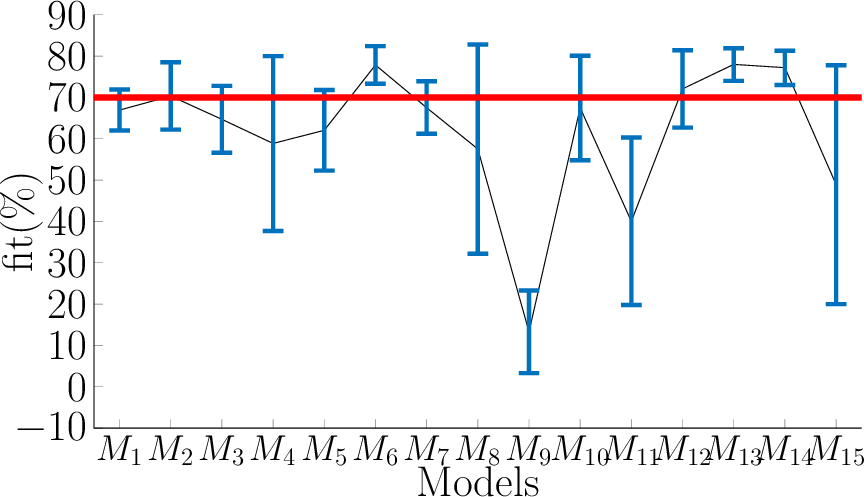} 
  \label{fig:pac2_val_20_min_31feat}
 }
 \subfloat[]{
  \includegraphics[width=0.33\textwidth]{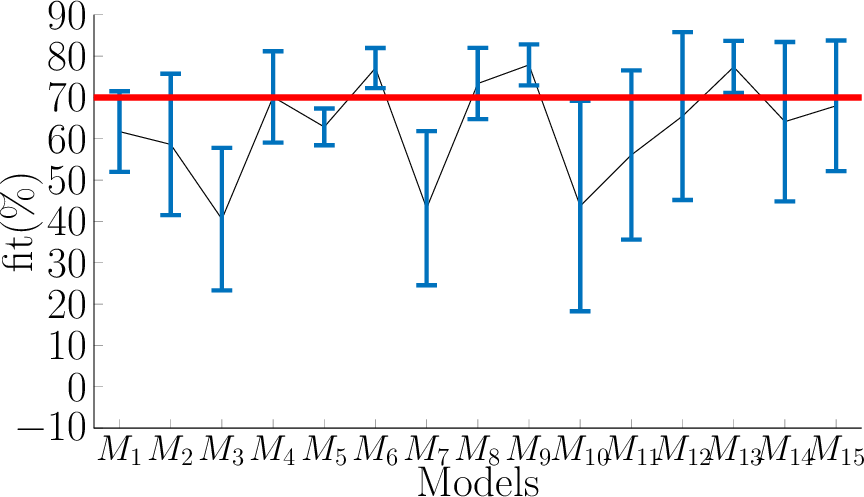} 
  \label{fig:pac2_val_20_min_32feat}
 }
 \\      
 \subfloat[]{
  \includegraphics[width=0.33\textwidth]{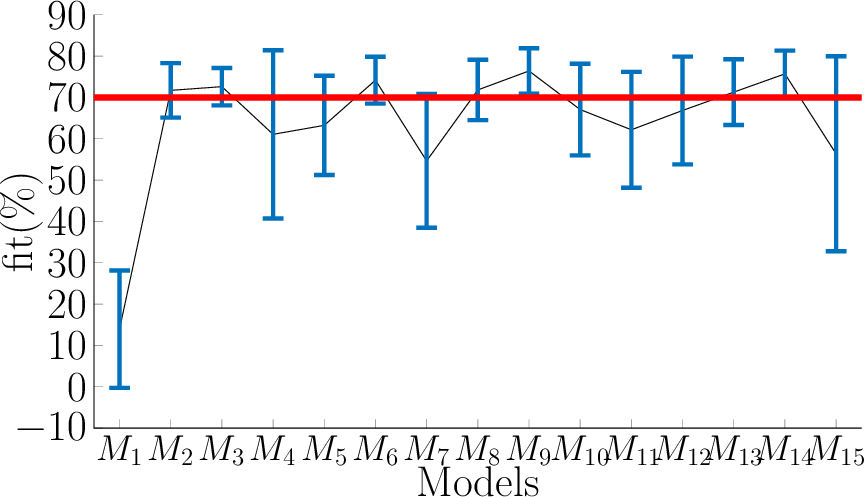} 
  \label{fig:pac2_val_20_min_33feat}
 }
 \subfloat[]{
  \includegraphics[width=0.33\textwidth]{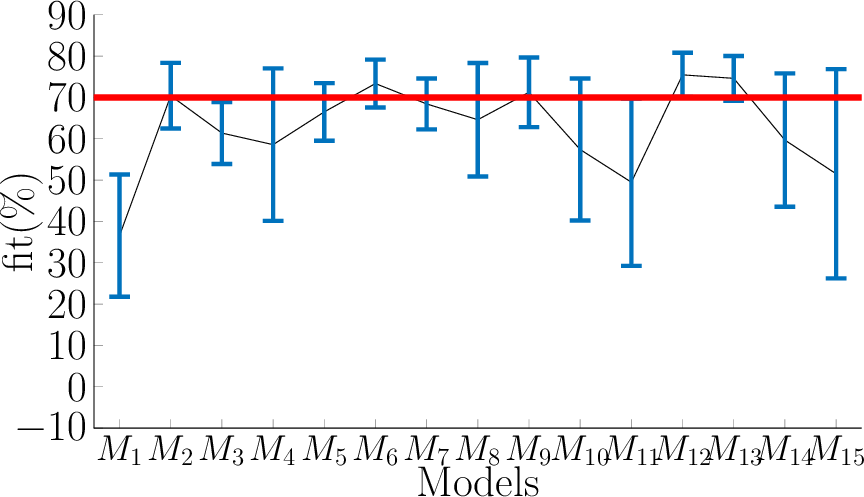} 
  \label{fig:pac2_val_20_min_34feat}
 } 
 \caption{Validation results for models from Patient A. 20 minutes
 forward horizon.  For each features set:({\bf a}) TEMP-EDA-HR-SpO2;
 ({\bf b}) TEMP-EDA-HR; ({\bf c}) TEMP-EDA-SpO2; ({\bf d})
 TEMP-HR-SpO2; ({\bf e}) EDA-HR-SpO2.}
\end{figure*}

\begin{figure*}
 \centering
 \subfloat[]{
  \includegraphics[width=0.33\textwidth]{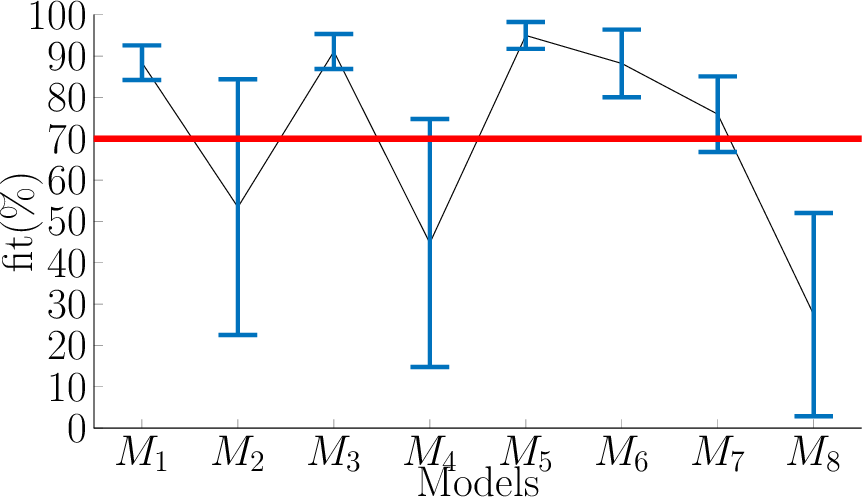} 
  \label{fig:pac11_val_20_min_4feat}
 }
 \subfloat[]{
  \includegraphics[width=0.33\textwidth]{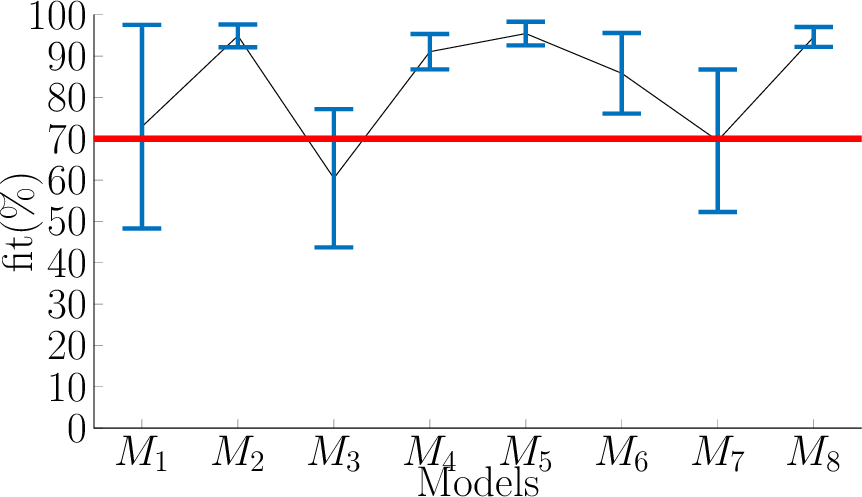} 
  \label{fig:pac11_val_20_min_31feat}
 }
 \subfloat[]{
  \includegraphics[width=0.33\textwidth]{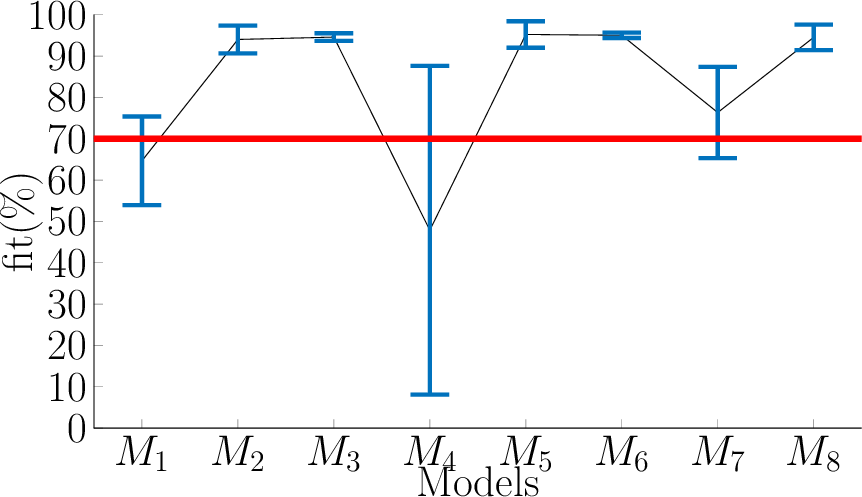} 
  \label{fig:pac11_val_20_min_32feat}
 }
 \\      
 \subfloat[]{
  \includegraphics[width=0.33\textwidth]{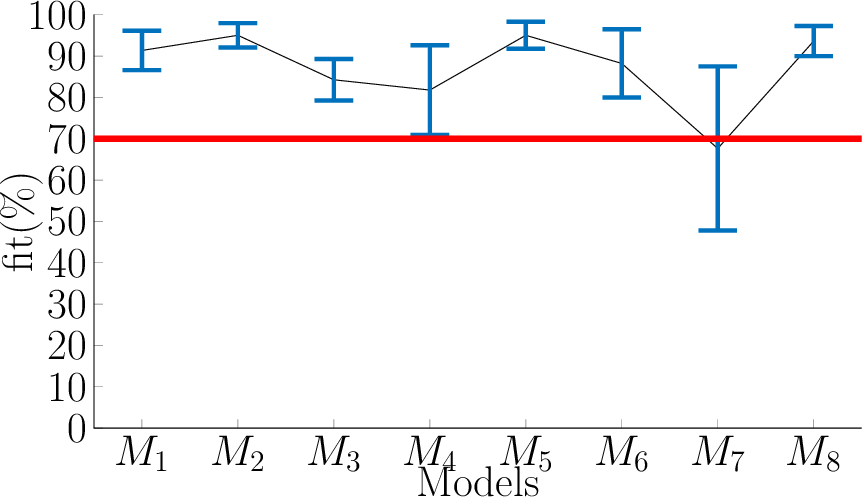} 
  \label{fig:pac11_val_20_min_33feat}
 }
 \subfloat[]{
  \includegraphics[width=0.33\textwidth]{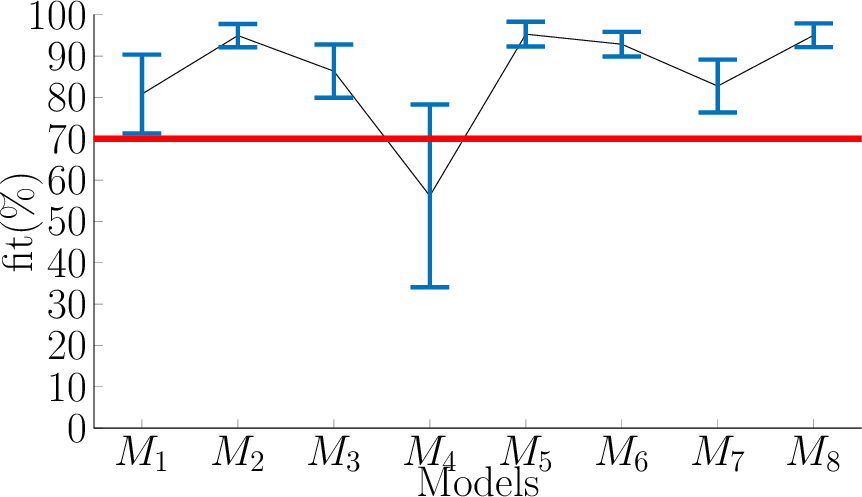} 
  \label{fig:pac11_val_20_min_34feat}
 } 
 \caption{Validation results for models from Patient B. 20 minutes
 forward horizon. For each features set:({\bf a}) TEMP-EDA-HR-SpO2;
 ({\bf b}) TEMP-EDA-HR; ({\bf c}) TEMP-EDA-SpO2; ({\bf d})
 TEMP-HR-SpO2; ({\bf e}) EDA-HR-SpO2.}
\end{figure*}

As the 20 minutes prediction horizon seems to be the safest horizon,
we use this to show the results for Patient A
in \figtoref{pac2_val_20_min_4feat}{pac2_val_20_min_34feat} and for
Patient B in
\figtoref{pac11_val_20_min_4feat}{pac11_val_20_min_34feat}. 
Horizontal axes in these figures represent each one of the validated
models. Vertical axes represent the average fit achieved, obtained as
the average of the $M-V_{ov}-1$ validations. The whiskers represent
the standard deviation, $\sigma$. $V_{ov}$, are the overfitted
validations (negative fit). These were removed to calculate the
average. The red line indicates the threshold set as fitting
criterion.

The deviations (the whiskers) for validations in Patient B
($\sigma_{B}$) are lower than deviations in Patient A
($\sigma_{A}$). This means that, the confidence of models from Patient
B should be higher than from Patient A. We can also state that these
models are more generalizable because the results for Patient B are
more consistent than those for Patient A, as data acquired from
Patient B have less discontinuities during monitorization.

Regarding the average values
in \figtoref{pac2_val_20_min_4feat}{pac2_val_20_min_34feat}, as the
four-features combination is a poor election, only 4 models have an
average fit higher that 70\%. For the TEMP-HR-SpO2 combination of
features (\figref{pac2_val_20_min_33feat}) we achieve the best
results. In this case, 7 models exceed the threshold of
70\%. Something similar occurs with the results for Patient B.

As aforementioned, to calculate the average fit, validations with
negative results have been removed. In some cases, the number of
useful validations is really low. This happens, for example, with the
validation of the model $M_{9}$ in \figref{pac2_val_20_min_31feat},
that only validates 3 migraines. The model $M_{1}$ for Patient B
validates also the same number of migraines
in \figref{pac11_val_20_min_4feat}, and only 2 migraines are validated
in \figref{pac11_val_20_min_31feat}
and \figref{pac11_val_20_min_32feat} (despite its high fitting).

As a result of the validation study: i) the four-features combination
is never the best option to predict migraines for any horizon length,
and ii) it seems that 20 minutes forward is the best window to predict
migraines for both patients. The first idea means that some biometric
variable worsens the prediction in combination with others (but not
itself). Hence, by removing one variable we achieve more useful
models. The second result achieves a prediction horizon close to the
constraints imposed by pharmacokinetics
(see \secref{introduction}). But, still, we pursue longer prediction
horizons and in next \subsec{improving} we show how to improve these
predictions.

\subsection{Improving predictions}
\label{subsec:improving}

This section is devoted to improve the prediction horizon. In this
section, the results of the schemes proposed
in \subsec{improving_meth} for the methodology are shown.

\begin{table*}
\caption{F value to compare the schemes proposed for the methodology}
        \centering
         \resizebox{\textwidth}{!}{
         \begin{tabular}{ccccccccccc}
         \toprule
         & \multicolumn{2}{c}{Base}
         & \multicolumn{2}{c}{Repair}
         & \multicolumn{2}{c}{FitGauss}
         & \multicolumn{2}{c}{Repair + FitGauss}
         & \multicolumn{2}{c}{FitGauss + Repair} \\ \midrule
         & Patient A & Patient B & Patient A & Patient B & Patient A &
         Patient B & Patient A & Patient B & Patient A & Patient B \\ \midrule
TPR (\%) & 24.7 & 30.4 & 29.8 & 36.3 & 31.0 & 39.0 & 33.1 & 39.6 & 30.6 & 39.0 \\ 
PPV (\%) & 45.1 & 60.7 & 57.2 & 90.4 & 75.1 & 71.6 & 80.5 & 81.1 & 78.5 & 87.9 \\ \midrule
F (\%) & 31.9 & 40.5 & 39.1 & 51.8 & 43.8 & 50.5 & 46.9 & 53.2 & 44.1 & 54.0 \\ 
         \bottomrule  
         \end{tabular}
         }
\label{tbl:F_value_schemes}
\end{table*}

We have studied all the repairing schemes during the validation stage
using the four-features combination (TEMP-EDA-HR-SpO2) and the 6
prediction horizons (from 10 to 60 minutes). The F value is used as
the metric to compare all the schemes with the basic one (results
in \subsec{validation}, scheme of \figref{dia_val_base}). To compute
the F value, the sensitivity (TPR) and the precision (PPV) values are
calculated. All values are based on the results of all the $M-1$
predictions of each $M_{i}$ model. This means that the true positive
(TP) account should be ideally $6*M*(M-1)$. The results are shown
in \tblref{F_value_schemes}.

\tblref{F_value_schemes} shows low levels of the F value because it has
been calculated as the average of the F values for every horizon. The
higher the horizon, the lower the F value, worsening this average F
value. The results show a high rate of false positives and low number
of detections for horizons higher than 40 minutes, as expected from
the training in \subsec{training}.

The best scheme for the proposed methodology is the combination of
repairing the prediction (remove spurious) and the Gaussian
fitting. The order (first repair then fitting) affects more to Patient
A than to Patient B. Therefore, the scheme Repair+fitGauss
in \figref{dia_val_all} is chosen as the repairing scheme of
symptomatic crises prediction.

\begin{figure*}[ht]
 \centering
 \subfloat[]{
  \includegraphics[width=0.5\textwidth]{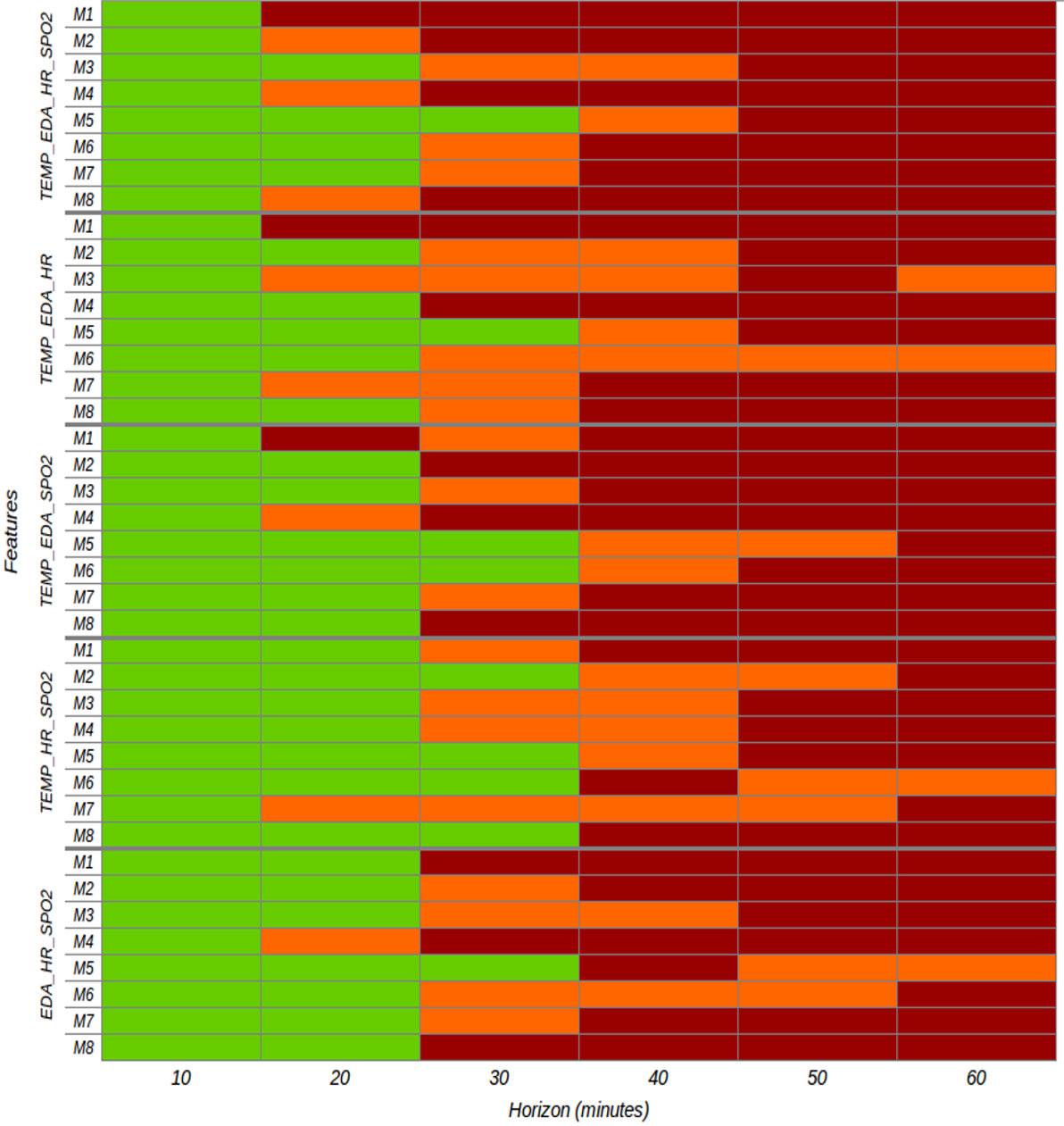} 
  \label{fig:pac11_val_colors_no_repair}
 }
 \subfloat[]{
  \includegraphics[width=0.5\textwidth]{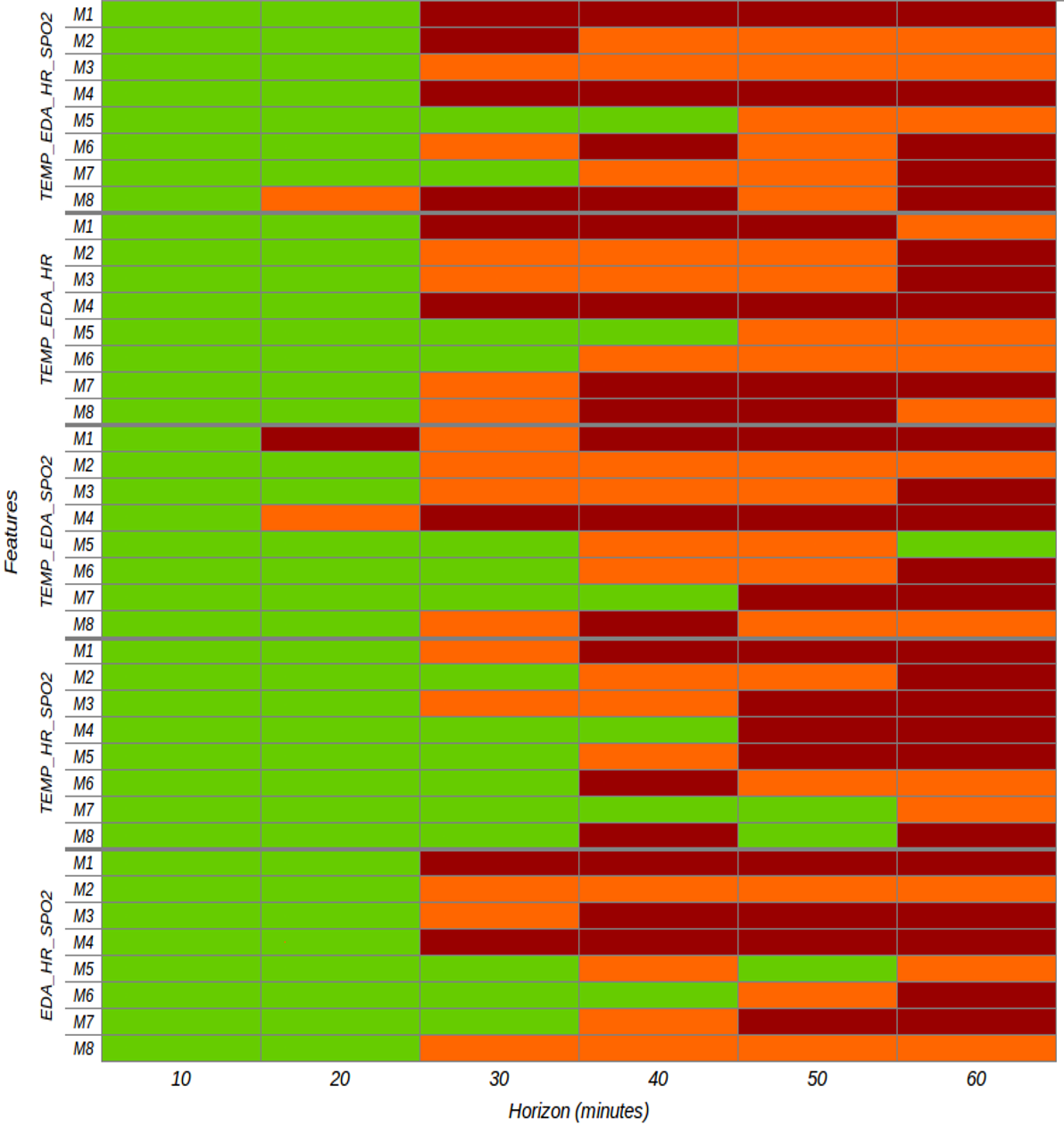} 
  \label{fig:pac11_val_colors_repair}
 } 
 \\
 \vspace{-0.25cm}        
 \captionsetup[subfigure]{labelformat=empty}
 \subfloat[]{
  \includegraphics[width=0.65\textwidth]{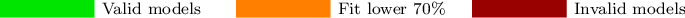} 
  \label{fig:colors_legend}
 }
 \vspace{-0.25cm}        

 \caption{Useful models in validation for Patient B at a 70\% of
 fit. ({\bf a}) Without reparation of prediction; ({\bf b}) After
 reparation of prediction and Gaussian fit.}

\end{figure*}

Now, the selected scheme is applied and compared with the base
scheme. For the sake of simplicity, only the results for Patient B are
presented in this section. \figref{pac11_val_colors_no_repair}
presents the results of validation for Patient B and all the trained
future horizons (10 to 60 minutes forward). Horizontal axis represents
the six different horizons trained and the vertical axis represents
the M = 8 models for each feature combination. Colors represent those
models good enough to be used as predictors in a real time
implementation (green), those with an average fit lower than $70\%$
(orange) and the discarded ones because the overfitting (red, less
than one third of the migraines available are validated).

All models validate all migraines for a prediction horizon of 10
minutes. As aforementioned, for 20 minutes forward almost all models
are useful, except model $M_{1}$ for some feature combinations. From
30 minutes ahead, there are not enough useful models, except in the
TEMP-HR-SpO2 feature combination, where 4 models validate quite
well. For prediction horizons equal and greater than 40 minutes,
migraine prediction is not possible, as pointed out in \subsec{training}
and \subsec{validation}.

As was introduced in \subsubsec{validation}, applying repairing
techniques to the prediction can increase the prediction horizon. In
this case we have applied reparation of the prediction and Gaussian
fitting, in this order. This is proved
by \figref{pac11_val_colors_repair}, again for Patient B. The average
prediction horizon has been incremented in 10 minutes (compared
to \tblref{num_models_average_all_horizons}), and some models validate
migraines with a future horizon equal to 40 minutes. There are
improvements in models for most of the prediction horizons and all
combination of features. These increments are due to removing false
positive detections, negative values, and values higher than the
maximum, 100, in the normalized symptomatic pain curve.

Regarding results for Patient A, the improvements achieved have
been lower. Although some more models are useful for 20 minutes, no
one is useful for 30 minutes of prediction if 70\% of fit is expected
(validating, at least, $\lceil M/3\rceil$ of the symptomatic crisis in
the training dataset).

As shown, the prediction horizon can be improved applying repairing
techniques to the predicted signal, reaching prediction times close to
the current time of pharmacokinetics and even exceeding
it. Additionally, we have shown a method to test the limits of a given
predictive model, that must be applied for each patient
individually. In particular, we have found that the maximum prediction
horizon for Patient A is in the interval $[20,30]$ minutes, and in the
interval of $[30,40]$ minutes for Patient B, the same results as
in~\cite{pagan2015robust}.

\begin{figure*}[ht]
 \subfloat[]{
  \includegraphics[width=1.0\textwidth]{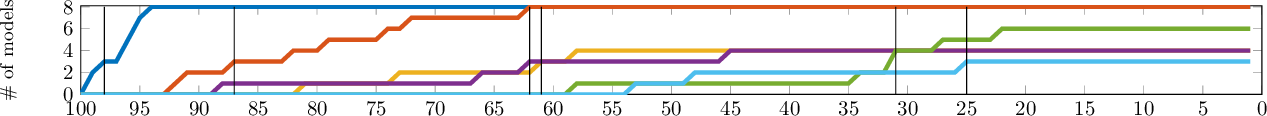} 
  \label{fig:strategies_model_selection_4vars}
 }
 \vspace{-0.45cm}        
\\
 \subfloat[]{
  \includegraphics[width=1.0\textwidth]{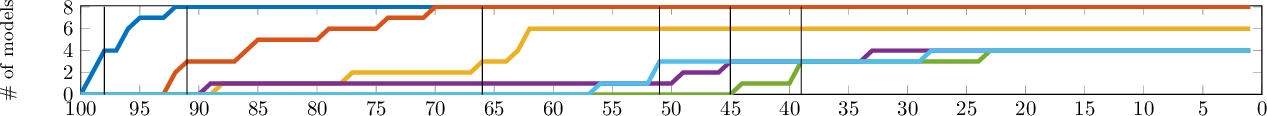} 
  \label{fig:strategies_model_selection_31vars}
 }
 \vspace{-0.45cm}        
\\
 \subfloat[]{
  \includegraphics[width=1.0\textwidth]{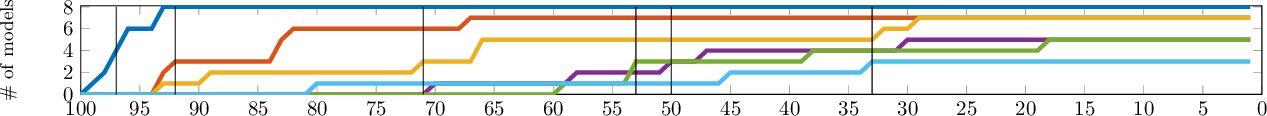} 
  \label{fig:strategies_model_selection_32vars}
 }
 \vspace{-0.45cm}        
\\
 \subfloat[]{
  \includegraphics[width=1.0\textwidth]{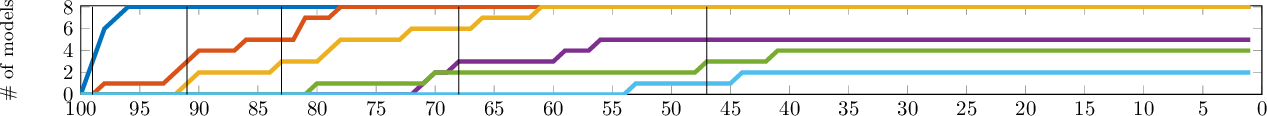} 
  \label{fig:strategies_model_selection_33vars}
 }
 \vspace{-0.45cm}        
 \\
 \subfloat[]{
  \includegraphics[width=1.0\textwidth]{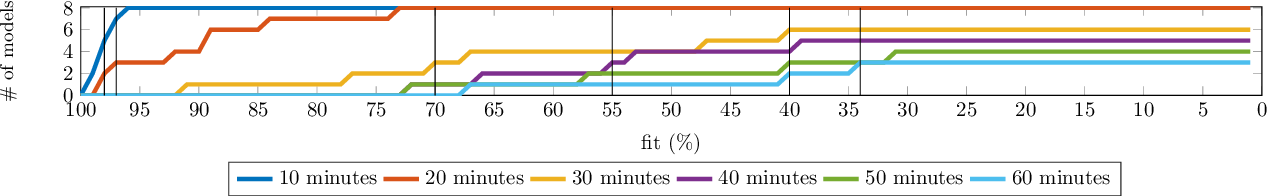} 
  \label{fig:strategies_model_selection_34vars}
 }
 \caption{Board of strategies for model selection for Patient
  B. Results after removing the spurious and applying the Gaussian fitting. ({\bf
  a}) TEMP-EDA-HR-SpO2; ({\bf b}) TEMP-EDA-HR; ({\bf c})
  TEMP-EDA-SpO2; ({\bf d}) TEMP-HR-SpO2; ({\bf e}) EDA-HR-SpO2}
 \label{fig:strategy_model_selection}
\end{figure*}

\subsection{Strategy for the selection of models}
\label{subsec:strategy}

For the results that have been presented, the fit value has been
prioritized, setting a fixed value of 70\% and observing the achieved
prediction horizon. As mentioned in \subsec{criteria}, in this section
we present two different strategies to select the models: i) regarding
the fit, or ii) regarding the prediction horizon. In all the cases we
always maintain that a model is considered good when it is able to
validate at least $\lceil M/3\rceil$ of the migraines from the dataset
at a given fit. The number of migraines to validate are 3 for Patient
B (8 migraines available in the training dataset) and 5 for Patient A
(15 migraines available in the training dataset). To avoid
overfitting and to calculate the average prediction, we still consider
as good the selection of, at least, $\lceil M/3\rceil$ migraines for
each feature combination.

\figref{strategy_model_selection} shows the number of models available
for every horizon at a desired filt level. As a reference, the
vertical bars mark the fit where $\lceil M/3\rceil$ models can predict
at least $\lceil M/3\rceil$ of the migraines in the training
dataset. If we focus more on the prediction, i) the first strategy
works setting a desired horizon and looking for the best feature
combination that reaches the maximum fit. On the contrary, if we focus
more on the fit level, ii) the second strategy works setting
a desired fit and looking for the feature combination that
reaches the farthest horizon. This is a more conservative selection,
for which we set the desired fit or goodness of the prediction and we
settle down the available horizon.

For instance, regarding \figref{strategy_model_selection}, if we are
looking for the best prediction horizon 20 minutes forward, we should
use the models calculated with EDA-HR-SpO2
in \figref{strategies_model_selection_34vars}, because the second
vertical bar has a higher fit for this feature combination than for
the others. There, we find 3 models validating at least 3 migraines
each one, in average, with a 97\% of fit. But if a prediction of 30
minutes forward is desired, the best option is to select the models
using the TEMP-HR-SpO2 feature combination
in \figref{strategies_model_selection_33vars}. For 50 minutes forward
we will select the models using
TEMP-EDA-SpO2 \figref{strategies_model_selection_32vars}, and accept
only a 50\% of fit.

\figref{strategy_model_selection} can also be used to look for a desired 
fit. For example, if the HR sensor is not available
(\figref{strategies_model_selection_32vars}) and we look for
predictions with a fit equal 60\% or higher, we could only satisfy a
horizon of 30 minutes. On the other hand, if the EDA sensor is not
available (\figref{strategies_model_selection_33vars}), for the same
minimum fit of 60\% we might predict up to 40 minutes.

It is worth noting that \figref{pac11_val_colors_repair} is just a
representation of the \figref{strategy_model_selection} at 70\% of
fit. In this figure, we can calculate at 1\% of fit how many models
that are not able to validate more than 3 migraines still remain.

This methodology leads to a versatile tool for the improvement of
predictions and the selection of models for predictions of symptomatic
crises in ambulatory real environments. This methodology has been
applied in a real clinical study of a disease with high
socio-economical impact. The effectiveness of the solution is studied,
and the results have proved to meet the pharmacokinetics limits
required to avoid the negative effects of symptomatic crises. The
results also show that for Patient A the limits of predictions are
between 20 and 30 minutes, and between 30 and 40 minutes for Patient
B, achieving fits of 70\% in both cases.

\subsection{Test results}
\label{subsec:test}
In this section, some test results are shown. Tests have been run
using the average model. This is the average of the prediction given
by the best $\lceil M/3\rceil$ models for each feature
combination. Over each prediction, as well as over the average of
these, the selected improvement scheme has been applied: spurious
removal and Gaussian fitting. The test dataset used is: 5 and 4
migraines, and 5 and 6 asymptomatic periods of time for Patients A and
B respectively.

A summary of the results is shown in \tblref{F_value_test}. As
expected, for Patient B, the results of the F value follow the trend
of the vertical bars in \figref{strategy_model_selection}, and the
best results are achieved for the feature combination
TEMP-HR-SpO2. Best results for Patient A are achieved for the feature
combination TEMP-EDA-SpO2. Besides, the worst results are achieved,
for both patients, with the combinations of four features. This leads
to conclude that the best model selection depends on: i) the features
used, ii) the desired horizon or iii) the desired fit.

Our results have been calculated using only data from two patients;
therefore, any generalization of the clinical conclusions obtained by
this study could be risky. However, the presented methodology, aim of
this work, can be validated by these results. In addition, it has been
shown that an analysis of the prediction horizon is needed in order to
improve the accuracy of the results, supporting our initial hypothesis.

\begin{table*}[ht]
\caption{Test results for Patients A and Patient B at 20 and 30 minutes of prediction
horizon respectively at 70\% of fit.}
        \centering
         \resizebox{\textwidth}{!}{
         \begin{tabular}{ccccccccccc}
         \toprule
         & \multicolumn{2}{c}{TEMP-EDA-HR-SpO2}
         & \multicolumn{2}{c}{TEMP-EDA-HR}
         & \multicolumn{2}{c}{TEMP-EDA-SpO2}
         & \multicolumn{2}{c}{TEMP-HR-SpO2}
         & \multicolumn{2}{c}{EDA-HR-SpO2} \\ \midrule
         & Patient A & Patient B & Patient A & Patient B & Patient A &
         Patient B & Patient A & Patient B & Patient A & Patient B \\ \midrule
TPR (\%) & 50.0 & 90.0 & 80.0 & 100 & 100 & 70.0 & 70.0 & 100 & 90.0 & 40.0 \\ 
PPV (\%) & 100 & 57.1 & 100 & 90.0 & 100 & 70.0 & 100 & 100 & 90.0 & 40.0 \\ \midrule
F (\%) & 66.7 & 47.1 & 88.9 & 90.0 & 100 & 77.8 & 82.4 & 100 & 90.0 & 47.1 \\ 
         \bottomrule  
         \end{tabular}
         }
\label{tbl:F_value_test}
\end{table*}

\begin{figure*}[ht]
 \centering
 \subfloat[]{ 
 \includegraphics[width=0.33\textwidth]{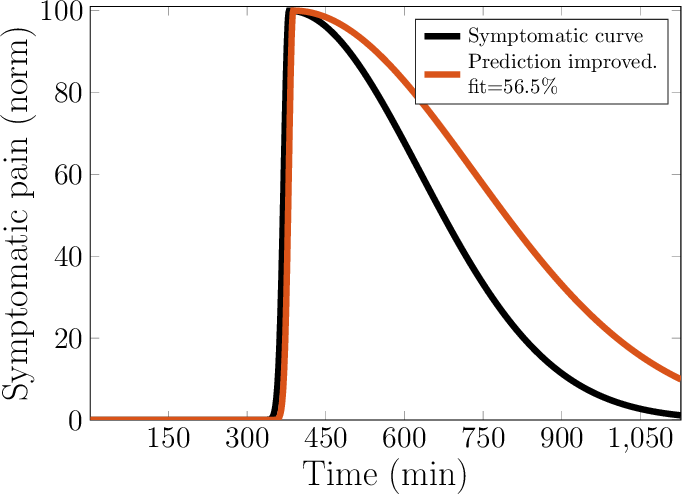} 
 \label{fig:test_pac2_m33_20min_TEMP_EDA_HR_SPO2}
 } 
 \subfloat[]{ 
 \includegraphics[width=0.33\textwidth]{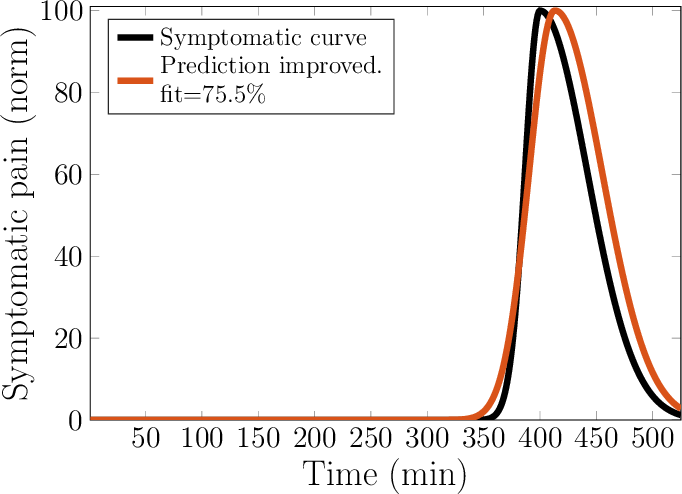} 
 \label{fig:test_pac2_m21_20min_EDA_HR_SPO2}
 } 
 \subfloat[]{ 
 \includegraphics[width=0.33\textwidth]{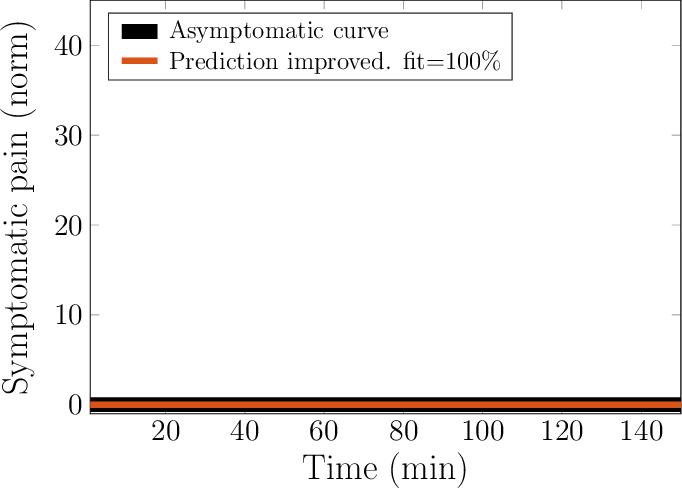} 
 \label{fig:test_pac2_M44_a1}
 } 
 \\ 
 \subfloat[]{ 
 \includegraphics[width=0.33\textwidth]{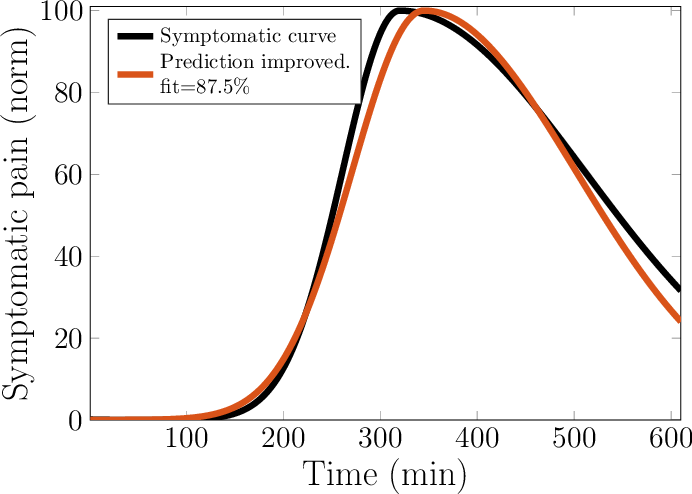} 
 \label{fig:test_pac11_m42_30min_TEMP_EDA_SPO2}
 } 
 \subfloat[]{ 
 \includegraphics[width=0.33\textwidth]{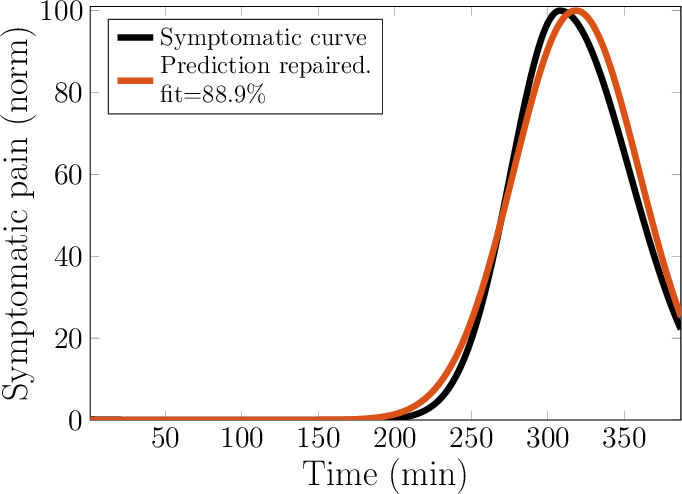}   
 \label{fig:test_pac11_m44_30min_TEMP_HR_SPO2}
 }
 \subfloat[]{ 
 \includegraphics[width=0.33\textwidth]{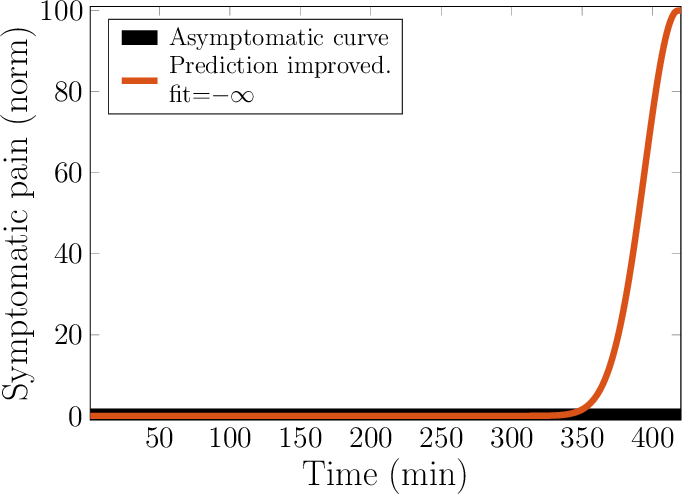}
 \label{fig:test_pac11_a3_30min_TEMP_EDA_HR_SPO2}
 } 
 \caption{Test results for symptomatic and asymptomatic periods. ({\bf
 a}) Patient A, TEMP-EDA-HR-SpO2, 20 min forward; ({\bf b}) Patient A,
 EDA-HR-SpO2, 20 min forward; ({\bf c}) Patient A, TEMP-EDA-SpO2 in an
 asymptomatic period, 20 min forward; ({\bf d}) Patient B,
 TEMP-EDA-SpO2, 30 min forward; ({\bf e}) Patient B, TEMP-HR-SpO2, 30
 min forward; ({\bf f}) Patient B, TEMP-EDA-HR-SpO2 in an
 asymptomatic period, 30 min forward.}
\end{figure*}

\figtoref{test_pac2_m33_20min_TEMP_EDA_HR_SPO2}{test_pac2_M44_a1} shows some
test results for Patient A
and \figtoref{test_pac11_m42_30min_TEMP_EDA_SPO2}{test_pac11_a3_30min_TEMP_EDA_HR_SPO2}
for Patient B. Several models applied over different feature sets are
presented to show the accuracy of the trained models.

For all the graphs, i) black curves represent the original symptomatic
curve that must be predicted, ii) the orange curves are the final
result after the reparation of the prediction and the Gaussian
fitting.

When a migraine occurs, models provide a prediction of some
symptomatic pain levels hours before the pain starts. Nevertheless,
these are false positive predictions, and the repairing process
removes them. The same happens with negative predictions or those
levels higher than 100. For all cases, repairing the prediction and
applying a Gaussian fitting leads us to improve the
fit. \figsref{test_pac2_M44_a1}{test_pac11_a3_30min_TEMP_EDA_HR_SPO2}
represent asymptomatic periods of time. The latest present a false
positive event not removed, obtained from prediction using the
TEMP-EDA-HR-SpO2 feature combination, that presents a high false
positive rate. The usage of the board of strategies to select the best
models (\figref{strategy_model_selection}) would have avoided these
false positives.


\section{Conclusions}
\label{sec:conclusions}

The experiments in this paper demonstrate that the use of state-space
models to predict symptomatic crises in chronic diseases is
time-limited. A methodology is presented as a versatile tool to
improve the quality of predictions of these crises, as well as to
increase the prediction horizon. This methodology selects models
according to the availability of sensors, and according to a desired
criteria of fit or prediction horizon. In this work we show how the
prediction time window of the disease can be calculated and that it
strongly depends on each patient. To prove our methodology, we present
a case study for migraine patients. Migraine models have been trained
up to 60 minutes in steps of 10 minutes, and it has been demonstrated
that state-space algorithms in combination with other techniques
(GPML, reparation of prediction and Gaussian fitting of the repaired
prediction) are currently limited up to 40 minutes to predict the
symptomatic crises of the migraine disease.

Migraines are one of the most disabling diseases, but we have shown
how migraine crises can be predicted using WBSNs in an ambulatory
way. The prediction horizons found are close or equal to 40
minutes---time enough to predict the migraine pain according to the
pharmacokinetics of current treatments---and much more accurate than
prodromic symptoms or auras.

Our methodology has proved to be capable of improving the prediction
horizon in a systematic way. Our results provide an effective
methodology for the selection of the future horizon in the development
of prediction algorithms for diseases experiencing symptomatic crises.


\section*{Acknowledgments}

Research by Josué Pagán has been funded by the Spanish Ministry of
Economy and Competitiveness under the Research Grant TEC2012-33892.

The authors also want to express their gratitude to the Service of
Neurology of Hospital Universitario La Princesa, whose help has been
precious for this work. In particular, doctors Ana Gago, Mónica
Sobrado, Mercedes Gallego and José A. Vivancos.


\section*{Conflicts of Interest}

The authors declare no conflict of interest.


\bibliographystyle{elsarticle-num}
\bibliography{n4sid-limits}

\end{document}